\documentclass[10pt,conference]{IEEEtran}

\IEEEoverridecommandlockouts
\IEEEpubid{\begin{minipage}{.83\textwidth}\ \\[30pt] \centering
    \copyright~2021~IEEE. Personal use of this material is permitted.
    Permission from IEEE must be obtained for all other uses, in any\break
    current or future media, including reprinting/republishing
    this material for advertising or promotional purposes, creating new\break
    collective works, for resale or redistribution to servers or lists, or
    reuse of any copyrighted component of this work in other works.
  \end{minipage}}

\usepackage[utf8]{inputenc}
\usepackage[T1]{fontenc}
\usepackage[english]{babel}

\usepackage{scalerel}
\usepackage{relsize}

\usepackage[bookmarks=false, pdfborder={0 0 0},hidelinks]{hyperref}

\usepackage{amsmath}
\usepackage{amssymb}
\usepackage{bm}
\usepackage{dsfont}
\usepackage{ulem}
\usepackage{nccmath}

\DeclareSymbolFont{largesymbolsA}{U}{txexa}{m}{n}
\DeclareMathSymbol{\varprod}{\mathop}{largesymbolsA}{16}

\usepackage{soul}

\usepackage[ruled]{algorithm2e}

\usepackage[capitalize, nameinlink]{cleveref}
\crefname{section}{Sec.}{Sec.}
\crefname{appendix}{Appx.}{Appx.}
\crefname{algorithm}{Alg.}{Alg.}
\crefname{figure}{Fig.}{Fig.}
\crefname{proposition}{Prop.}{Prop.}
\crefname{table}{Table}{Tables}
\crefname{definition}{Def.}{Def.}
\crefname{theorem}{Thm.}{Thm.}
\crefname{enumi}{}{}
\crefformat{footnote}{#2\footnotemark[#1]#3}

\usepackage{nameref}
\usepackage{csquotes}

\def\no/{\textcolor{darkred}{\faTimesCircle}}
\def\yes/{\textcolor{darkgreen}{\faCheckSquare}}

\usepackage[most]{tcolorbox}
\newtcolorbox{resbox}[1][]{
  colback=gray!35,left=0pt,top=0pt,right=0pt,bottom=0pt,
  boxrule=0pt, colframe=white,
  #1}
\usepackage{tabularx}
\usepackage{booktabs}
\usepackage[inline]{enumitem}
\usepackage{subcaption}
\usepackage{adjustbox}
\usepackage{threeparttable}
\usepackage{multirow}
\usepackage{proof}
\usepackage{pgfplotstable}
\usepackage{pgfplots}
\usepackage{tikz}
\usetikzlibrary{shapes, automata, arrows, intersections,
  fit, calc, positioning, tikzmark, patterns}
\usepackage{pifont}

\usepackage{listing}
\newcommand{\cinl}[1]{\lstinline[language=c]$#1$}
\newcommand{\asminl}[1]{\lstinline[language={[x86masm]Assembler}]$#1$}
\newcommand{\bashinl}[1]{\lstinline[language=bash]$#1$}

\usepackage{eso-pic}

\usepackage{fontawesome}

\renewcommand{\paragraph}[1]{\medskip \noindent {\bf #1.}}

\newcommand*{\origrightarrow}{}
\let\origrightarrow\textrightarrow
\renewcommand*{\textrightarrow}{\fontfamily{cmr}\selectfont\origrightarrow}
\newcommand*{\longleftrightarrows}{\hstretch{1.5}{\leftrightarrows}}

\usepackage{mdframed}

\definecolor{darkblue}{rgb}{0.3,0.3,1}
\definecolor{darkgreen}{rgb}{0, 0.5, 0}
\definecolor{darkred}{rgb}{0.72,0.04,0.04}

\definecolor{asmlab}{HTML}{a1a100}
\definecolor{asmpre}{HTML}{7d8f29}
\definecolor{asmkey}{HTML}{0000ff}
\definecolor{asmreg}{HTML}{19177d}
\definecolor{asmflg}{HTML}{bd7a00}
\definecolor{asmval}{HTML}{870000}
\definecolor{cstring}{HTML}{b72020}
\definecolor{ckey}{HTML}{008000}
\definecolor{ctok}{HTML}{666666}

\definecolor{gitblue}{HTML}{f1f8ff}
\definecolor{gitgreen}{HTML}{e6ffed}
\definecolor{gitred}{HTML}{ffeef0}
\definecolor{gitmidblue}{HTML}{dbedff}
\definecolor{gitmidgreen}{HTML}{cdffd8}
\definecolor{gitmidred}{HTML}{ffdce0}
\definecolor{gitdarkgreen}{HTML}{acf2bd}
\definecolor{gitdarkred}{HTML}{fdb8c0}

\colorlet{bggrey}{gray!02}

\newcommand{\mintsize}{\smaller[3]}

\newlength{\mlapf}
\newlength{\mlaps}

\newcommand{\mintflg}[1]{{\tt\textcolor{asmflg}{#1}}}
\newcommand{\mintreg}[1]{{\tt\textcolor{asmreg}{#1}}}
\newcommand{\mintval}[1]{{\tt\textcolor{asmval}{#1}}}
\newcommand{\mintkey}[1]{{\tt\textcolor{asmkey}{#1}}}

\newcommand{\minttok}[1]{{\tt\textcolor{ctok}{#1}}}

\makeatletter
\def\customlabel#1#2{%
  \begingroup
  #2%
  \def\@currentlabel{#2}%
  \phantomsection\label{#1}%
  \endgroup
}
\newrobustcmd\gliteral{%
  \begingroup
  \let\do\@makeother\dospecials
  \catcode`\{=1
  \catcode`\}=2
  \glit@inline
}%
\newcommand\glit@inline[1]{%
  \endgroup
  \textcolor{lightgray}{\textquoteleft}%
  \textbf{\lstinline[language=c]{#1}}%
  \textcolor{lightgray}{\textquoteright}
}
\makeatother

\newcommand{\gtoken}[2][]{$\langle$\textit{#2}#1$\rangle$}
\newcommand{\goption}[1]{[ #1 ]}

\def\good/{\textcolor{darkgreen}{\checkmark}}
\def\poor/{\textcolor{orange}{\checkmark}}
\def\var/{\textcolor{orange}{$\thicksim$}}
\def\bad/{\textcolor{darkred}{$\bm\times$}}
\def\na/{{\sc N/a}}

\def\goto/{\textcolor{asmkey}{goto }}

\newcommand{\fmacro}[3]{\texttt{\textcolor{#1}{#2\textsuperscript{#3}}}}
\newcommand{\ftype}[2][]{\fmacro{asmreg}{#2}{#1}}
\newcommand{\fname}[2][]{\fmacro{asmval}{#2}{#1}}
\newcommand{\ffunc}[2][]{\textbf{\fmacro{ckey}{#2}{#1}}}
\newcommand{\fstruct}[1]{\fmacro{asmkey}{#1}{}}
\newcommand{\fsubst}[2]{#1\ffunc{<}#2\ffunc{>}}
\newcommand{\fget}[2]{#1(#2)}

\def\fset/{\fstruct{set}}
\def\fbool/{\ftype{bool}}

\def\faddress/{\fname{A}}

\def\fcode/{\fname{C}}
\def\fcodet/{\fname[$\lozenge$]{C}}
\def\fmachine/{\fname{M}}

\def\ft/{\fname{t}}
\def\fe/{\fname{e}}
\def\fl/{\fname{l}}
\def\fr/{\fname{r}}

\def\fasm/{\ftype{asm}}
\def\fmstate/{\ftype{mstate}}
\def\flocation/{\ftype{location}}
\def\fregister/{\ftype{register}}
\def\fvalue/{\ftype{value}}
\def\fexpression/{\ftype{expression}}

\def\feval/{\ffunc{eval}}
\def\fexec/{\ffunc{exec}}

\def\ftoken/{\ftype{token}}
\def\fT/{\fname{T}}
\def\fasmt/{\ftype[$\lozenge$]{asm}}
\def\fI/{\fname{I}}
\def\fbo/{\fname[O]{B}}
\def\fbi/{\fname[I]{B}}
\def\fst/{\fname[T]{S}}
\def\fsc/{\fname[C]{S}}
\def\fmbarrier/{\fname{F}}

\def\fasme/{\fname{X}}

\def\flocationt/{\ftype[$\circ$]{location}}
\def\fsmash/{\ffunc{may\_impact}}
\def\fdomain/{\ffunc{domain}}
\def\fdomains/{\ffunc{domain$^{*}$}}
\def\fsmashs/{\ffunc{impact$^{*}$}}
\def\flocations/{\ftype[$*$]{location}}
\def\fM/{\fname{M}}

\newcommand\fequiv[3]{$\smash{\stackrel{_#1}{\sim}}_{#3}^{#2}$}
\newcommand\fequivbis[3]{$\smash{\stackrel{_#1}{\cong}}_{#3}^{#2}$}

\def\fequivt/{\fequiv{\lozenge}{\fT/}{\fname{B}}}
\def\fequivtt/{\fequiv{\lozenge}{\fT/_1, \fT/_2}{\fname{B}}}
\def\fequivm/{\fequiv{\bullet}{}{}}
\def\fequivtm/{\fequivbis{\blacklozenge}{\fT/}{B, \fmbarrier/}}
\def\fequivO/{\fequivbis{\blacklozenge}{\fT/}{\fbo/, \fmbarrier/}}
\def\fequivI/{\fequivbis{\blacklozenge}{\fT/}{\fbi/, \fmbarrier/}}
\def\fequivOT/{\fequivbis{\blacklozenge}{\fT/_1, \fT/_2}{\fbo/, \fmbarrier/}}
\def\fequivIT/{\fequivbis{\blacklozenge}{\fT/_1, \fT/_2}{\fbi/, \fmbarrier/}}

\def\abi/{ABI}

\def\gnu/{GNU}
\def\llvm/{LLVM}

\def\caml/{OCaml}

\def\gcc/{GCC}
\def\gccversion/{10.2}
\def\clang/{Clang}
\def\icc/{ICC}
\def\compcert/{CompCert}
\def\visualstudio/{Visual Studio}

\def\gas/{gas}

\def\alsa/{ALSA}
\def\ffmpeg/{ffmpeg}
\def\gmp/{GMP}
\def\libyuv/{libyuv}
\def\libatomic/{libatomic\_obs}
\def\libtomcrypt/{libtomcrypt}
\def\haproxy/{haproxy}
\def\xfstt/{xfstt}
\def\dropbear/{dropbear}
\def\udpcast/{UDPCast}
\def\px264/{x264}
\def\glibc/{glibc}

\def\x86/{x86}
\newcommand{\reg}[1]{\mintreg{\%#1}}
\def\al/{\reg{al}}
\def\eax/{\reg{eax}}
\def\edx/{\reg{edx}}
\def\cl/{\reg{cl}}
\def\ecx/{\reg{ecx}}
\def\ebx/{\reg{ebx}}
\def\edi/{\reg{edi}}
\def\esi/{\reg{esi}}
\def\ebp/{\reg{ebp}}
\def\esp/{\reg{esp}}

\def\arm/{ARM}

\def\riscv/{RISC-V}

\def\binsec/{\textsc{Binsec}}
\def\framac/{\mbox{Frama-C}}

\def\tina/{\textsc{TInA}}
\def\proto/{\textsc{RUS}\tina/}
\def\Proto/{Repair User Specification by \tina/}

\def\nbchunks/{2656}        
\def\nbpackages/{202}
\def\nbfaultypackages/{85}
\def\nbstrongfaultypackages/{54}
\def\nbissues/{2183}      
\def\nbstrongissues/{986}   
\def\nbpatchissues/{2000}
\def\nbpatchstrongissues/{803}
\def\patchratio/{92}
\def\patchratiosig/{81}

\def\nbsubmittedpatchs/{114}     
\def\nbacceptedpatchs/{38}

\def\nbsubmittedsolvedissues/{538}     
\def\nbacceptedsolvedissues/{156}

\def\nbsubmittedsolvedprojects/{8}
\def\nbacceptedsolvedprojects/{7}

\def\rqcheck/{\cref{item:rq1}}
\def\rqbugs/{\cref{item:rq2}}
\def\rqpatch/{\cref{item:rq3}}
\def\rqimpact/{\cref{item:rq4}}
\def\rqinternal/{\cref{item:rq5}}
\def\rqrefine/{\cref{item:rq6}}

\def\Cpp/{\mbox{C${+}{+}$}}

\title{Interface Compliance of Inline Assembly: \\ Automatically Check, Patch and Refine}

\def\ceaA/{Univ. Paris-Saclay, CEA, List\\Saclay, France}
\def\verimagA/{Univ. Grenoble Alpes, VERIMAG\\Grenoble, France}

\author{
  \IEEEauthorblockN{Frédéric Recoules}
  \IEEEauthorblockA{
    \ceaA/
    \\\href{mailto:frederic.recoules@cea.fr}
    {frederic.recoules@cea.fr}}
  \medskip
  \IEEEauthorblockN{Matthieu Lemerre}
  \IEEEauthorblockA{
    \ceaA/
    \\\href{mailto:matthieu.lemerre@cea.fr}
    {matthieu.lemerre@cea.fr}}
  \and
  \IEEEauthorblockN{Sébastien Bardin}
  \IEEEauthorblockA{
    \ceaA/
    \\\href{mailto:sebastien.bardin@cea.fr}
    {sebastien.bardin@cea.fr}}
  \medskip
  \IEEEauthorblockN{Laurent Mounier}
  \IEEEauthorblockA{
    \verimagA/
    \\\href{mailto:laurent.mounier@univ-grenoble-alpes.fr}
    {laurent.mounier@univ-grenoble-alpes.fr}}
  \and
  \IEEEauthorblockN{Richard Bonichon}
  \IEEEauthorblockA{Tweag I/O
    \\Paris, France
    \\\href{mailto:richard.bonichon@gmail.com}
    {richard.bonichon@gmail.com}}
  \medskip
  \IEEEauthorblockN{Marie-Laure Potet}
  \IEEEauthorblockA{
    \verimagA/
    \\\href{mailto:marie-laure.potet@univ-grenoble-alpes.fr}
    {marie-laure.potet@univ-grenoble-alpes.fr}}
}

\begin{document}

\maketitle

\begin{abstract}
  Inline assembly is still a common practice in low-level C programming, typically for efficiency reasons or for accessing specific hardware resources.
Such embedded assembly codes in the GNU syntax (supported by major
compilers such as \gcc/, \clang/  and \icc/) have an {\it interface} specifying
how the assembly codes interact with the C environment.  For simplicity reasons, the compiler treats GNU inline  assembly codes as  blackboxes
and relies only on their interface to correctly glue them into the compiled C code.
Therefore, the adequacy between the assembly chunk and its interface (named {\it compliance}) is of primary importance, as such compliance issues
can lead to subtle and hard-to-find bugs.
We propose  \proto/, the first automated technique for formally  checking inline assembly compliance, with the extra ability to propose (proven) patches
and (optimization) refinements in certain cases.
\proto/ is based on an original formalization of the inline assembly compliance problem together with novel dedicated algorithms.
Our prototype has been evaluated on \nbpackages/  Debian packages with inline assembly (\nbchunks/ chunks), finding \nbissues/ issues  in \nbfaultypackages/
packages
-- \nbstrongissues/ significant issues in \nbstrongfaultypackages/ packages (including  major projects such as \ffmpeg/ or \alsa/),
and proposing patches for \patchratio/\% of them. Currently,
\nbacceptedpatchs/ patches have already been accepted
(solving \nbacceptedsolvedissues/ significant issues),  with positive feedback from development teams.
\end{abstract}


\section{Introduction}

\vspace*{-.2em}

\paragraph{Context}
 Inline assembly, i.e.~embedding assembly code inside a higher-level host language,
is still a common practice in low-level C/\Cpp/ programming, for efficiency reasons or for accessing specific hardware resources --
it is typically widespread  in  resource-sensitive areas such as cryptography, multimedia,
  drivers, system, automated trading or video games~\cite{8952223,Rigger:2018:AXI:3186411.3186418}.
 Recoules et al.~\cite{8952223} estimate that 11\% of Debian
 packages written in C/\Cpp/ directly or indirectly depend on
 inline assembly, including major projects such as \gmp/ or \ffmpeg/,
 while 28\% of the top rated C projects on GitHub contain
 inline assembly according to Rigger et al.~\cite{Rigger:2018:AXI:3186411.3186418}.

Thus, compilers supply a syntax to  embed assembly
instructions in the source program. The most widespread is the {\it GNU inline assembly syntax},
driven by \gcc/ but also supported by  \clang/ or
\icc/.
The GNU syntax provides an {\it interface} specifying
how the assembly code interacts with the C environment. The compiler then treats GNU inline  assembly codes as  blackboxes
and relies only on this interface to correctly insert them into the compiled C
code\footnote{Microsoft inline assembly is different and has no interface, see ~\cref{sec:masm}.}.

\paragraph{Problem}
The problem with GNU inline assembly is twofold.
First, it is hard to write correctly\footnote{From the \textit{llvm-dev} mailing list~\cite{1_STANNARD_2018}: \textquote{\gcc/-style inline assembly is notoriously hard to write correctly}.}:
 inline assembly syntax~\cite{gccdoc} is not beginner-friendly, the language
 itself is neither standardized nor fully described,
 and some corner cases are defined by  \gcc/ implementation
 (with occasional changes from time to time).
Second, assembly chunks are treated as blackboxes, so that the compiler does not
do any sanity checks%
\footnote{Note that syntactically incorrect assembly
  instructions are caught during the translation from assembly to  machine code.
  \vspace{.49cm}\phantom{.}} and {\it assumes} the embedded assembly code respects its
declared interface.

Hence, in addition to  usual  functional bugs  in the assembly instructions themselves, inline assembly is also prone  to {\it interface compliance} bugs, i.e.,~mismatches between
the declared interface and the real behavior of the assembly  chunk which can lead to subtle and hard-to-find bugs -- typically incorrect results or crashes due to
either subsequent  compiler optimizations or ill-chosen register allocation.
In the end,  compliance issues can lead to severe bugs (segfault, deadlocks, etc.) and, as they depend on low-level compiler choices,  they are hard to identify and can hide
for years before being triggered by a compiler update.
For example,  a  2005 compliance bug introduced in the {\tt \libatomic/}
  library of lock-free primitives for multithreading
 made deadlocks possible:
 it was identified and patched only in 2010 (commit \href{https://github.com/ivmai/libatomic_ops/commit/03e48c173c0a46abf8b3272b43f9c2e4f4f4170e}{03e48c1}).
 A similar bug was still lurking in another primitive in 2020 until we automatically found and patched it (commit \href{https://github.com/ivmai/libatomic_ops/commit/d728ce4e2be5c8328f0af8fc738622915c520aee}{05812c2}).
We also found a 1997 interface compliance bug in \glibc/ (leading to a
\textsf{segfault} in a  string primitive) that was patched in 1999 (commit \href{https://github.com/bminor/glibc/commit/7c97addd6fbb44818b6e4d219cdbd189554a10f3}{7c97add}),
then reintroduced in 2002 after  refactoring.

\paragraph{Goal and challenges}
\textit{We address the challenge of helping developers write safer inline assembly  code
  by designing and developing  automated  techniques helping to achieve interface compliance, i.e.~ensuring that both
  the assembly template and its interface are consistent with each other}.
This is challenging for several reasons:
\begin{description}
\item [Define] The method must be built on a (currently missing) proper   formalization of interface compliance,  both
  realistic and amenable to automated formal verification;

\item [Check, Patch \& Refine] The method must be able to check whe\-ther an assembly chunk complies with its interface, but ideally it should also be able to
  automatically suggest patches for bugs or code refinements;
\item [Wide applicability] The method must be generic enough to encompass several architectures, at least \x86/ and \arm/.

\end{description}

Fehnker et al.~\cite{Fehnker08} published the only attempt we know of
 to inspect  the interface written by the developer. Yet,
their definition of interface compliance is syntactic and incomplete
-- for example they cannot detect the \glibc/ issue mentioned above.
Moreover, they do not cover all subtleties of \gcc/ inline assembly (e.g., token constraints),  consider only compliance checking (neither patching nor refinement) and the
implementation is tightly bound to \arm/ (much simpler than \x86/).

Note that recent attempts for verifying codes mixing C and assembly
\cite{8952223,DBLP:conf/uss/CorteggianiCF18}
simply \textit{assume} interface compliance.

\paragraph{Proposal and contributions}
We propose \proto/,  the first {\it sound} technique 
for comprehensive automated interface compliance checking, automated
patch synthesis and interface refinements.  We claim  the following contributions:
\begin{itemize}

\item a novel {\it semantic} and {\it comprehensive} formalization of the problem of interface compliance (\cref{sec:formal}), amenable to
  formal verification;

\item a new {\it semantic} method (\cref{sec:rustina})  to  
 automatically verify
   the compliance of inline assembly chunks,
  to generate a corrective patch for the majority of compliance issues
   and additionally to suggest interface refinements;

\item thorough experiments (\cref{sec:evaluation}) of a prototype implementation (\cref{sec:implementation})
  on a large set of \x86/ real-world examples (all inline assembly found in a Debian Linux distribution)  demonstrate that \proto/ is able to automatically check and curate
a large code base (202 packages, 2640 assembly chunks) in {\it a few minutes}, detecting 2036 issues
and solving 95\% of them;

\item a 
study of  current inline assembly coding practices  (\cref{sec:assessment});
  besides  identifying the common \textit{compliance} issues found in the wild (\cref{sec:xp:check}),
  we also exhibit 6 recurring  patterns leading to the vast majority (97\%) of compliance issues and show that
  5 of them rely on {\it fragile} assumptions and
  can lead to serious bugs  (\cref{sec:assessment}).

\end{itemize}

As the time of writing, \nbacceptedpatchs/
patches have already been accepted by \nbacceptedsolvedprojects/ projects,
 solving \nbacceptedsolvedissues/ significant issues (\cref{sec:xp:impact}).

\paragraph{Summary}
Inline assembly is
a delicate practice.
\proto/ aids developers in achieving interface compliant
inline assembly code.
  Compliant assembly chunks can still be buggy but \proto/
  {\it automatically}  removes  a whole class of problems.
Our technique has  already helped several renowned projects fix code, with positive feedback from  developers.

\smallskip

{\it \noindent Note: supplementary material, including prototype and
  benchmark data, is available online~\cite{rustina_in_a_nutshell_2020}.}

\section{Context and motivation}
\label{sec:motivation}

\begin{figure}[htbp]
  \centering
  \begin{subfigure}[b]{.41\textwidth}
    \includegraphics{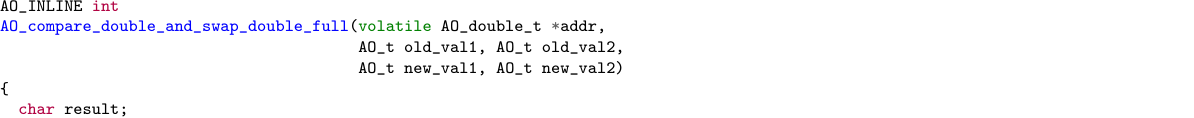}
    {\vspace{-1ex} \mintsize \quad \textelp{}}
    \\~\\\vphantom{.}
    \includegraphics{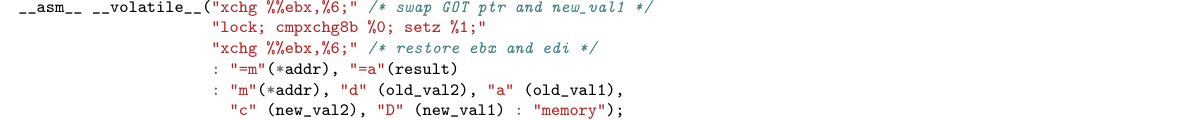}
    {\vspace{-1ex} \mintsize \quad \textelp{}}
    \\~\\\vphantom{.}
    \includegraphics{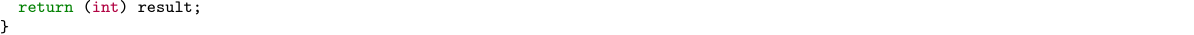}
  \end{subfigure}
  \caption{\bashinl{atomic_ops/sysdeps/gcc/x86.h}@\href{https://github.com/ivmai/libatomic_ops/commit/30cea1b9ea06c4c25cc219e1197dfac8dfa52083}{30cea1b}}
  \label{fig:motiv_c_src}
  \label{fig:form_c_src}
\end{figure}

The code in \cref{fig:motiv_c_src} is an extract from \libatomic/, commit \href{https://github.com/ivmai/libatomic_ops/commit/30cea1b9ea06c4c25cc219e1197dfac8dfa52083}{30cea1b} dating back to early 2012.
It was replaced 6 months later by commit
\href{https://github.com/ivmai/libatomic_ops/commit/64d81cd475b07c8a01b91a3be25e20eeca2d27ec}{64d81cd}
because it led to a \texttt{segmentation fault}
when compiled with \clang/.
By 2020, another \textit{latent} bug was still lurking
\textit{until automatically discovered and patched by our prototype}
\proto/
(commit \href{https://github.com/ivmai/libatomic_ops/commit/d728ce4e2be5c8328f0af8fc738622915c520aee}{05812c2}).

\paragraph{What the code is about}
This function uses inline assembly to implement the standard atomic primitive
\textit{Compare And Swap} -- i.e. write \cinl{new_val} in \cinl{*addr}
if this latter still equals to \cinl{old_val}
(where 8-byte values \cinl{old_val} and \cinl{new_val}    are split in 4-byte values \cinl{old_val1}, \cinl{old_val2}, etc.).
The assembly statement (syntax discussed in \cref{sec:syntax}) comprises
assembly instructions (e.g., \cinl{"lock; cmpxchg8b \%0;"})
building an \textit{assembly template} where some operands have been replaced by
\textit{tokens} (e.g., \cinl{\%0}) that will be latter assigned by the compiler. It also has a specification, the
\textit{interface}, binding together assembly registers, tokens and C
expressions:
line 196 declares the {\it outputs}, i.e.~C values expected to be
assigned by the chunk; lines 197 and 198 declare the {\it inputs}, i.e.~
C values the compiler should pass to the chunk. The string placed before
a C expression is called a {\it constraint} and indicates the set of
possible assembly operands this expression can be bound to by the compiler.
For instance, \cinl{"d" (old_val2)} indicates that register \edx/
should be initialized with the value of \cinl{old_val2}, while \cinl{"=a" (result)}
indicates the value of \cinl{result} should be collected from \eax/.
Token \cinl{\%0}
introduced by \cinl{"m" (*addr)} is an indirect memory access: its address,
arbitrarily denoted \cinl{\&0} here, can be bound to several possibilities
(cf.~\cref{fig:x86_domains}) -- including \esi/ or \ebx/.

\Cref{fig:motiv_ir} gives the functional meaning of this binding
 along with the semantics of
the assembly instructions (where ``$::$'' is the concatenation,
``\smash{$\overset{c}{\longleftarrow}$}'' a conditional assignment,
``$e_{\{h..l\}}$'' the bits extraction and ``\textsf{\smaller zext}$_{n}$'' the zero extension to size $n$).

\begin{figure}[htbp]
  \centering
  \newlength{\minheight}
  \setlength{\minheight}{\totalheightof{\ttfamily\scriptsize ()fLlthdkgjp}}
  \begin{tikzpicture}[
      asm/.style={minimum height=\minheight,
        /utils/exec={\ttfamily}, font=\scriptsize},
      block/.style={fit=#1, draw, inner ysep=0pt, inner xsep=5pt,
        rounded corners=1pt},
      ]

      \node[asm] (d) {$\longleftarrow$};
      \node[asm, anchor=east] (dl) at (d.west) {\textcolor{asmflg}{$z$}};
      \node[asm, anchor=west] (dr) at (d.east)
      {(\edx/ $::$ \eax/) = \cinl{*(\&0)}};

      \node[asm, anchor=north] (e) at (d.south) {$\longleftarrow$};
      \node[asm, anchor=east] (el) at (e.west) {\edx/ $::$ \eax/};
      \node[asm, anchor=west] (er) at (e.east) {\cinl{*(\&0)}};

      \node[asm, anchor=north] (f) at (e.south)
      {\smash{\makebox[\widthof{$\longleftarrow$}]{
            $\overset{\textcolor{asmflg}{z}}{\longleftarrow}$}}
        \vphantom{$\longleftarrow$}};
      \node[asm, anchor=east] (fl) at (f.west) {\cinl{*(\&0)}};
      \node[asm, anchor=west] (fr) at (f.east) {\ecx/ $::$ \ebx/};

      \node[block=(dl)(d)(dr)(el)(e)(er)(fl)(f)(fr), asmpre] (j) {};
      \node[anchor=south east] at (j.south east) {\textcolor{asmpre}{\faLock}};

      \path let \p1 = (f), \p2 = (j.south) in
      node[asm, anchor=north] (g) at (\x1, \y2) {$\longleftarrow$};
      \node[asm, anchor=east] (gl) at (g.west) {\eax/};
      \node[asm, anchor=west] (gr) at (g.east)
      {\eax/$_{\{31..8\}}$ $::$ (\textsf{\smaller zext}$_{8}$ \textcolor{asmflg}{$z$})};

      \node[asm, anchor=north] (h) at (g.south) {$\longleftrightarrows$};
      \node[asm, anchor=east] (hl) at (h.west) {\ebx/};
      \node[asm, anchor=west] (hr) at (h.east) {\edi/};

      \node[asm, anchor=north] (i) at (h.south) {$\longleftarrow$};
      \node[asm, anchor=east] (il) at (i.west) {\cinl{result}};
      \node[asm, anchor=west] (ir) at (i.east) {\eax/$_{\{7 .. 0\}}$};

      \path let \p1 = (d), \p2 = (j.north) in
      node[asm, anchor=south] (c) at (\x1, \y2) {$\longleftrightarrows$};
      \node[asm, anchor=east] (cl) at (c.west) {\ebx/};
      \node[asm, anchor=west] (cr) at (c.east) {\edi/};

      \node[asm, anchor=south] (w) at (c.north) {$\longleftarrow$};
      \node[asm, anchor=east] (wl) at (w.west) {\edi/};
      \node[asm, anchor=west] (wr) at (w.east) {\cinl{new_val1}};

      \node[asm, anchor=south] (x) at (w.north) {$\longleftarrow$};
      \node[asm, anchor=east] (xl) at (x.west) {\ecx/};
      \node[asm, anchor=west] (xr) at (x.east) {\cinl{new_val2}};

      \node[asm, anchor=south] (y) at (x.north) {$\longleftarrow$};
      \node[asm, anchor=east] (yl) at (y.west) {\eax/};
      \node[asm, anchor=west] (yr) at (y.east) {\cinl{old_val1}};

      \node[asm, anchor=south] (z) at (y.north) {$\longleftarrow$};
      \node[asm, anchor=east] (zl) at (z.west) {\edx/};
      \node[asm, anchor=west] (zr) at (z.east) {\cinl{old_val2}};

      \node[asm, anchor=south] (a) at (z.north) {$\longleftarrow$};
      \node[asm, anchor=east] (al) at (a.west) {\cinl{\&0}};
      \node[asm, anchor=west] (ar) at (a.east) {\cinl{addr}};

      \path let \p1 = ($(j.west) + (left:.4)$), \p2 = (e) in node (l)
      at (\x1, \y2)
      {\phantom{$\longrightarrow$}};
      \node[asm, anchor=east, gray] (m) at (l.west)
      {\textcolor{asmkey}{cmpxchg8b} \cinl{\%0}};

      \path let \p1 = (m.west), \p2 = (c) in
      node[asm, anchor=west] at (\x1, \y2)
      {\textcolor{asmkey}{xchg} \ebx/, \edi/};

      \path let \p1 = (m.west), \p2 = (d) in
      node[asm, anchor=west, gray] (n) at (\x1, \y2)
      {\textcolor{asmpre}{lock}};

      \path let \p1 = (m.west), \p2 = (g) in
      node[asm, anchor=west] (o) at (\x1, \y2)
      {\textcolor{asmkey}{setz} \al/};

      \path let \p1 = (m.west), \p2 = (h) in
      node[asm, anchor=west] at (\x1, \y2)
      {\textcolor{asmkey}{xchg} \ebx/, \edi/};

      \path let \p1 = (m.west), \p2 = (i) in
      node[asm, anchor=west] at (\x1, \y2)
      {\cinl{"=a" (result)}};

      \path let \p1 = (m.west), \p2 = (w) in
      node[asm, anchor=west] at (\x1, \y2)
      {\cinl{"D" (new_val1)}};

      \path let \p1 = (m.west), \p2 = (x) in
      node[asm, anchor=west] at (\x1, \y2)
      {\cinl{"c" (new_val2)}};

      \path let \p1 = (m.west), \p2 = (y) in
      node[asm, anchor=west] at (\x1, \y2)
      {\cinl{"a" (old_val1)}};

      \path let \p1 = (m.west), \p2 = (z) in
      node[asm, anchor=west] at (\x1, \y2)
      {\cinl{"d" (old_val2)}};

      \path let \p1 = (m.west), \p2 = (a) in
      node[asm, anchor=west] at (\x1, \y2)
      {\cinl{"=m" (*addr)}};

  \end{tikzpicture}
  \caption{Assembly statement semantics}
  \label{fig:motiv_ir}
  \label{fig:formal_code}
\end{figure}

This example allows us to introduce the concept of interface compliance issues
and the associated miscompilation problems:
\begin{enumerate*}[label={\Alph*)}]
\item \label{elt:frame}
 {\bf (framing condition)} {\it incomplete} interfaces,   possibly leading to  miscompilations due to  wrong data dependencies;
\item \label{elt:unicity}
 {\bf (unicity)} {\it ambiguous}  interfaces, where the result  depends on compiler choices for token allocation.
\end{enumerate*}

\paragraph{\cref{elt:frame} An incomplete frame definition}
Here, register \edx/ is declared as
\textit{read-only} (by default, \textit{non-output} locations are) whereas it
is overwritten  by
instruction \mintkey{cmpxchg8b} (c.f. \cref{fig:motiv_ir}).
\edx/ should be declared as output as well.

{\it Impact:} The compiler exclusively relies on the interface to know the
\textit{framing-condition} -- i.e. which locations are read or written.
When this information is incomplete, data dependencies are miscalculated,
potentially leading to incorrect optimizations. Here, the compiler  believes
\edx/  still contains \cinl{old_val2} after the
assembly chunk is executed, while it is not the case.

Note that  \ebx/ and \esi/ are {\it not} missing the output
  attribute:  while overwritten by the \mintkey{xchg} instructions,
  they are then  restored to their initial value.

\paragraph{\cref{elt:unicity} Ambiguous interface}
Here, while most of the binding is fixed, the compiler still has to bind
\cinl{\&0} according to  constraint \cinl{"m"}.
Yet, if the compiler rightfully chooses \ebx/,
the 
data dependencies in the assembly itself
differ from the expected one:  pointer \cinl{addr} is exchanged with
\cinl{new_val1} just before being dereferenced, which is not the expected behaviour.
The problem here is that the result cannot be predicted as it depends on
token resolution  from the compiler.

{\it Impact:} the function is likely to end up in a \texttt{segmentation fault} when compiled by \clang/.
Historically, \gcc/ was not able to select \ebx/ and the bug did not manifest,
but \clang/ did not had such restriction.

\paragraph{The problem}  These compliance issues are really hard to find out either manually or syntactically.
First,  there is here clearly no hint from the assembly template itself
(\cinl{"cmpxchg8b \%0"}) that register \edx/ is modified.
Second, complex token binding and aliasing constraints  must be taken into account.
Third, subtle data flows must be taken into account -- for example a read-only value
modified then restored is not a compliance issue.

\paragraph{\proto/ insights}
To circumvent these problems, we have developed \proto/, an automated tool
to check inline assembly compliance (i.e. formally verifying the absence of
compliance errors) and to patch the  identified issues.

\proto/ builds upon an original formalization of the inline assembly interface  compliance problem,
{\it encompassing both framing  and unicity}. From that, our method lifts
binary-level Intermediate Representation (sketched in \cref{fig:motiv_ir}) and adapt the classical data-flow
analysis framework (\textit{kill-gen}~\cite{10.1145/512927.512945}) in order to achieve
sound interface compliance verification -- especially  \proto/  reasons about
token assignments.
 From the expected interface, it infers for each token an
overapproximation of the set of valid locations 
and then computes  the set of locations that shall not be altered
before the token is used.
Here, it deduces that writing in  register \ebx/ may
impact  token
\cinl{\%0}.
Also, it detects that a write  occurs in the read-only
register \edx/, thus  successfully reporting the two  issues.

\begin{figure}[htbp]

  \centering
  \includegraphics{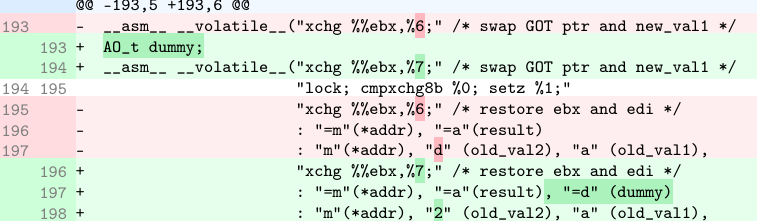}

\caption{Frame-write corrective patch}
\label{fig:motiv_patch}
\end{figure}

Moreover,
\proto/ 
automatically suggests %
  patches for the two issues.
For framing,
\Cref{fig:motiv_patch} highlights the core differences %
between the two versions (\edx/ is now rightfully declared as output with
\cinl{"=d"}) -- a similar patch now lives on the
current version of the function
(commit \href{https://github.com/ivmai/libatomic_ops/commit/d728ce4e2be5c8328f0af8fc738622915c520aee}{05812c2}).
For unicity, it
suggests to declare \ebx/ as clobber, yielding a working fix.
Yet, it also over-constrains the interface --
the syntax does not allow a simple disequality between \cinl{\%0} and \ebx/.
Developers actually patched the issue in 2012 in a completely different way by
 rewriting the assembly template
(commit \href{https://github.com/ivmai/libatomic_ops/commit/64d81cd475b07c8a01b91a3be25e20eeca2d27ec}{64d81cd}) --
such a solution is out of \proto/'s scope.

Generic and automatic, our approach is well suited to handle what expert
developers failed to detect, while a simpler ``bad'' patterns detection approach
would struggle against both  the combinatorial complexity induced by the size of  architecture instruction
sets and the underlying reasoning complexity (dataflow, token assignments).
Overall, \proto/ found and patched many other significant issues
in several well-known  open source projects (\cref{sec:xps}).

\section{GNU inline assembly syntax}
\label{sec:syntax}
\paragraph{Overview}
This feature allows the insertion of assembly instructions anywhere
in the code without the need to call an externally defined function.
\cref{fig:grammar} shows the concrete syntax of
an inline assembly block, which can either be
\textbf{basic} when it contains only the assembly template or
\textbf{extended} when it is supplemented by an \textit{interface}.
This section concerns the latter only. The assembly statement
consists of \textquote{a series of low-level instructions that convert
  input parameters to output parameters}~\cite{gccdoc}. The
\textit{interface} binds C lvalues (i.e., expressions evaluating to C memory locations) and expressions to assembly
operands specified as {\it input} or {\it output},   and declares a list of {\it clobbered} locations (i.e., registers or
memory cells whose values could change).
For the sake of completeness, the statement can also be tagged with \mintval{volatile},
\mintval{inline} or
\mintval{goto} qualifiers, which are irrelevant for interface compliance, thus not discussed in this paper.
The interface bindings described above are written by \mintval{string}
specifications, which we will now explain.

\paragraph{Templates}
The assembly text is given in the form of a \textit{formatted}
\mintval{string} template that, like \cinl{printf}, may contain so-called \textit{token}s (i.e., place holders).
These start with \cinl{\%} followed by an optional \textit{modifier}
and a reference to an entry of the \textit{interface}, either by name (an
\textit{identifier} between square brackets) or by a number denoting a
positional argument. The compiler preprocesses the template, substituting
\textit{token}s by assembly operands according to the entries and the modifiers
(note that only  a subset of \x86/ modifiers is fully
documented~\cite{30527_gcc_bugzilla_2007}) and then emits it \textit{as is} in
the assembly output file.

\begin{figure}[htbp]
  \scriptsize
  \setlength{\tabcolsep}{2pt}
  \begin{tabular}{rcl}
    \gtoken{statement} & ::= & \gliteral{asm} \goption{\gliteral{volatile}}
                               \gliteral{(} \gtoken[:\mintval{string}]{template}
                               \goption{\gtoken{interface}} \gliteral{)} \\
    \gtoken{interface} & ::= & \gliteral{:}
                               \goption{\gtoken{outputs}}
                               {\gliteral{:}
                               \goption{\gtoken{inputs}}
                               {\gliteral{:}
                               \goption{\gtoken{clobbers}}}} \\

    \addlinespace
    \gtoken{outputs}   & ::= & \gtoken{output}
                               \goption{\gliteral{,} \gtoken{outputs}} \\
    \gtoken{inputs}    & ::= & \gtoken{input}
                               \goption{\gliteral{,} \gtoken{inputs}} \\
    \gtoken{clobbers}  & ::= & \gtoken[:\mintval{string}]{clobber}
                               \goption{\gliteral{,} \gtoken{clobbers}} \\
    \addlinespace
    \gtoken{output}    & ::= & \goption{\gliteral{[} \gtoken{identifier}
                               \gliteral{]}}
                               \gtoken[:\mintval{string}]{constraint}
                               \gliteral{(} \gtoken{Clvalue} \gliteral{)} \\
    \gtoken{input}    & ::= & \goption{\gliteral{[} \gtoken{identifier}
                               \gliteral{]}}
                               \gtoken[:\mintval{string}]{constraint}
                               \gliteral{(} \gtoken{Cexpression} \gliteral{)} \\
  \end{tabular}
  \caption{Concrete syntax of an extended assembly chunk}
  \label{fig:grammar}
\end{figure}

\paragraph{Clobbers}
They are names of hard registers whose values may be modified by the
execution of the statement, but not intended as output.
Clobbers must not overlap with inputs and outputs.
The \cinl{"cc"} keyword identifies, when it exists, the conditional flags
register.
The \cinl{"memory"} keyword instructs the compiler that arbitrary
memory could be accessed or modified.
\begin{figure}[!htbp] \footnotesize
  \begin{align*}
    \text{\mintval{a}} &= \{ \text{\eax/} \}
    & \text{\mintval{b}} &= \{ \text{\ebx/} \}
    & \text{\mintval{c}} &= \{ \text{\ecx/} \}
    \\
    \text{\mintval{d}} &= \{ \text{\edx/} \}
    & \text{\mintval{S}} &= \{ \text{\esi/} \}
    & \text{\mintval{D}} &= \{ \text{\edi/} \}
  \end{align*}\vspace{-1.5em}%
  \begin{align*}
    \text{\mintval{U}} &= \text{\mintval{a}} \cup \text{\mintval{c}}
                         \cup \text{\mintval{d}}
    & \text{\mintval{q}} = \text{\mintval{Q}}
    &= \text{\mintval{a}} \cup \text{\mintval{b}}
      \cup \text{\mintval{c}} \cup \text{\mintval{d}}
      \\
    \text{\mintval{i}} &= \text{\mintval{n}}
    = \mathds{Z}
    &
    \text{\mintval{r}} = \text{\mintval{R}}
    &= \text{\mintval{q}} \cup \text{\mintval{S}}
      \cup \text{\mintval{D}} \cup \{ \text{\ebp/} \}
  \end{align*}\vspace{-1.5em}%
  \begin{align*}
    \text{\mintval{p}} = \{ r_{b} \bm+ k \bm\times r_{i} \bm+ c
    &\text{ for } r_{b} \in \text{\mintval{r}} \cup
      \{ \text{\esp/} \} \cup \{ 0 \}
    \\
    &\text{ and } r_{i} \in \text{\mintval{r}} \cup \{ 0 \}
    \text{ and } k \in \{ 1, 2, 4, 8 \} \\
    &\text{ and } c \in \text{\mintval{i}} \}
  \end{align*}\vspace{-1.5em}%
  \begin{align*}
    \text{\mintval{m}} &= \{ \cinl{*}p \text{ for }
                         p \in \text{\mintval{p}} \}
    & \text{\mintval{g}} &= \text{\mintval{i}}
                           \cup \text{\mintval{r}}
                           \cup \text{\mintval{m}}
  \end{align*}
  \vspace{-2em}
  \caption{\gcc/ i386 architecture constraints   }
  \label{fig:x86_domains}
\end{figure}

\paragraph{Constraints}
A third language describes the  set of valid
assembly operands for token assignment. The latter are of 3 kinds:
an  immediate
  value,
a register 
or a memory  location.
\cref{fig:x86_domains} gives a view  of
common {\bf atomic constraints} (``letters'')  used in \x86/.
Constraint entries can have more that one atomic constraint (e.g., \cinl{"rm"}),
in which case the compiler chooses among the {\bf union} of operand choices.
The language allows to organize constraints into
{\bf multiple alternatives}, separated by \gliteral{,}.
Additionally,
\begin{enumerate*}
\item [\textbf{matching constraint}] between an input token and an output token forces them to be equal;

\item [\textbf{early clobber}] \gliteral{&}
  informs the compiler that
   it must not attempt to use the same operand for this
  output and any non-matched input;

\item [\textbf{commutative pair}] \gliteral{
    makes an input and the next one exchangeable.
\end{enumerate*}

Finally, output constraints must start either with \gliteral{=} for the write-only
mode or with \gliteral{+} for the read-write permission.

\section{Formalizing Interface Compliance}
\label{sec:formal}

\subsection{Extended assembly}

\paragraph{Assembly chunks}
We denote by \fcode/: \fasm/  a standard chunk of assembly code. Such a chunk
operates over a memory state \fmachine/: \fmstate/, that is a map from
\flocation/ (registers of the underlying architecture or memory cells) to
basic values \value/ (int8, int16, int32, etc.).
We call \faddress/: \fvalue/
\fset/ the set  of valid addresses for a given architecture.  The value of an
expression in a given memory state is given by  function \feval/: \fmstate/
$\times$ \fexpression/ $\mapsto$ \fvalue/.  The set of valid assembly
expressions is architecture-dependent (\cref{fig:x86_domains} is for i386). We
abstract it as a set of \fexpression/s built over registers, memory accesses
\cinl{*} and operations.  Finally, an assembly chunk \fcode/ can be executed in
a memory state \fmachine/ to yield a new memory state
\fmachine/$'$ %
  with  function \fexec/: \fasm/ $\times$ \fmstate/
$\mapsto$ \fmstate/.
\Cref{fig:asm_types} recaps above functions and types.

\begin{figure}[htbp]
  {
    \centering
    \smaller[1]
    \setlength{\tabcolsep}{2pt}
    \begin{tabular}{rcl}
      \fexec/
      & : & \fasm/ $\times$
            \fmstate/ $\mapsto$
            \fmstate/

      \\[0pt]

    \feval/
      & : & \fmstate/ $\times$
            \fexpression/ $\mapsto$
            \fvalue/

      \\[0pt]

      \fmstate/
      & : & \flocation/ $\mapsto$ \fvalue/

      \\[0pt]

      \fexpression/ \mintkey{as} $e$
      & ::= & \fvalue/  \textbf{|}  \fregister/ \textbf{|} \cinl{*}$e$
              \textbf{|} $e$ $+$ $e$
              \textbf{|} $e$ $\times$ $e$
              \textbf{|} ...

      \\[0pt]

      \flocation/
      & ::= & \fregister/
              \textbf{|} \fvalue/

      \\[0pt]

      \fregister/
      & ::= & \eax/ \textbf{|} \ebx/ \textbf{|}
              \ecx/ \textbf{|} \edx/ \textbf{|} ... // case of \x86/

      \\[0pt]

      \fvalue/
      & : & \cinl{int8}
            \textbf{|} \cinl{int16}
            \textbf{|} \cinl{int32}
            \textbf{|} ...
    \end{tabular}
  }
  \caption{Assembly types}
  \label{fig:asm_types}
\end{figure}

\paragraph{Assembly templates}
Inline assembly does not directly use assembly chunks, but rather \textit{assembly templates},
denoted \fcodet/: \fasmt/, which are assembly chunks where some operands are replaced by
so-called \textit{tokens}, i.e., placeholders for regular assembly \fexpression/s
to be filled by the compiler (formally, they are identifiers \cinl{\%0}, \cinl{\%1}, etc.).
Given a \textit{token assignment} %
\fT/:  \ftoken/ $\mapsto$ \fexpression/, we can turn  an assembly template \fcodet/:\fasmt/ into a regular assembly chunk  \fcode/:\fasm/ using standard syntactic substitution \ffunc{<>}, denoted 
 \fsubst{\fcodet/}{\fT/}:\fasm/.
The value of \ftoken/ \ft/ through  assignment \fT/ is given by
 \feval/(\fmachine/, \fget{\fT/}{\ft/}).

\paragraph{Formal interface}
We model an \textit{interface} \fI/ $\triangleq$ (\fbo/, \fbi/, \fst/, \fsc/,
\fmbarrier/) as a tuple consisting of    \textit{output tokens}  \fbo/:\ftoken/
\fset/, \textit{input tokens}\footnote{Actually, a
    concrete interface also contains initializer and collector expressions in
    order to bind I/O assembly locations input and output to C. We skip them
    for clarity, as they do not impact compliance.}  \fbi/:\ftoken/ \fset/ ,
a  \textit{memory separation} flag \fmbarrier/: \fbool/,  {\it clobber registers} \fsc/: \fregister/ \fset/
and {\it valid token assignments} \fst/: \fT/ \fset/.

\begin{itemize}

\item Input and output tokens bind the assembly memory state and the C
  environment. Informally, the locations pointed to by tokens in \fbi/ are {\it input} {
    initialized} by the value of some C expressions while the values of the
  tokens in \fbo/ are {\it output} to some C lvalues.  \fbo/ $\cup$
  \fbi/ contains all token declarations and  \fbo/ $\cap$
  \fbi/ may be non-empty;

\item If the flag \fmbarrier/ is set to false, then
  assembly instructions may have side-effects on the C environment -- otherwise they operate on separate memory parts;

\item \fsc/ and \fst/ provide additional information about how the
  compiler should instantiate the assembly template to machine code: the clobber
  registers in \fsc/ can be used for temporary computations during
  the execution (their value is possibly modified by the chunk), while
  \fst/ represents all possible token assignments the compiler is allowed to
  choose -- the GNU syntax typically leads to equality, disequality and
  membership constraints between tokens and (sets of) registers.
\end{itemize}

\paragraph{Extended assembly chunk} An extended assembly chunk \linebreak \fasme/ $\triangleq$ (\fcodet/, \fI/)
is a pair made of an assembly template \fcodet/ and its related interface \fI/.
The assembly template is the operational content of the chunk (modulo token assignment)
while the interface is a contract between the chunk, the C environment and
low-level location management.

\subsection{(Detail) From GNU concrete syntax to formal interfaces} \label{sec:translate}
Let us see how the formal interface \fI/  is derived from concrete \gnu/ syntax
(\cref{fig:grammar}).  Tokens \fbo/ and \fbi/ come from the corresponding
output and input lists except that:
\begin{enumerate*}[label=\alph*)]
\item if an output entry is declared using the
\cinl{'+'} modifier then it is added to both \fbo/ and \fbi/;  and
\item  if an input token and an output token are necessarily mapped to the same register,
  they are unified.
\end{enumerate*}
Each register in the clobber list
belong to \fsc/.  If the clobber list contains \cinl{"memory"}, the memory
separation flag \fmbarrier/ is  false, true otherwise.
The set \fst/ of
valid token assignments \fT/
is formally derived  in 3 steps:
 \begin{enumerate}
 \item  collection of string constraints, splitting constraints by alternative (i.e., \cinl{','}):
   \cinl{(}\ftoken/~$\mapsto$~\mintval{string}\cinl{)} \fset/;

 \item architecture-dependent (e.g., \cref{fig:x86_domains}) evaluation of string constraints:
   \cinl{(}\ftoken/~$\mapsto$~\fexpression/~%
   \fset/\cinl{)}~\fset/; representing a disjunction of  conjunctions of atomic membership constraints \ftoken/ $\in$   \{ \cinl{exp}, \ldots, \cinl{exp} \};
 \item flattening: (\ftoken/ $\mapsto$ \fexpression/) \fset/ representing a disjunction of conjunctions of atomic equality constraints  \ftoken/ =  \fexpression/;
 \end{enumerate}

\noindent
Still, token assignments must respect the following properties (and are filtered out otherwise):
\begin{itemize}
\item every output \ftoken/ maps to an assignable operand,
  either a \fregister/ or a \cinl{*}\fe/ \fexpression/;
\item every output \ftoken/ maps to distinct \flocation/;
\item each  \ftoken/ maps to a \textit{clobber-free} \fexpression/
\end{itemize}

\noindent
where  a {\it clobber-free} \fexpression/ is an expression without any
clobber register  nor any early-clobber sub-expression (i.e. containing
the mapping of an early-clobber \ftoken/, introduced by the \cinl{'\&'}
modifier).

\Cref{fig:formal_interface} exemplifies the interface formalization
of \cref{fig:form_c_src}'s chunk introduced in
\cref{sec:motivation}.
Tokens \fbo/ and \fbi/ simply enumerate the
present entries respectively in output and input lists (L196-198).  The
5\textsuperscript{th} entry matches the same register \eax/ as the second,
\cinl{\%4} is unified with \cinl{\%1}.
For the sake of brevity, we split the set of token assignments into two parts:
one invariant w.r.t. compiler choices, and one that may vary (we only list
4 of them but there are other valid combination of memory references).
Finally, it
has no clobbered register and, because of  keyword \cinl{"memory"},
memory separation is \cinl{false}.

\begin{figure}[htbp]
    \mintsize
    \setlength{\tabcolsep}{2pt}
    \begin{tabular}{ccll}
      \fbo/ & = & \{ \cinl{\%0}, \cinl{\%1} \}, \\
      \fbi/ & = & \{ \cinl{\%2}, \cinl{\%3}, \cinl{\%5}, \cinl{\%6} \} \\
      \fst/ & = &
                  \{
                   [
                  \cinl{\%1} $\mapsto$ \eax/,
                  \cinl{\%3} $\mapsto$ \edx/,
                  \cinl{\%5} $\mapsto$ \ecx/,
                  \cinl{\%6} $\mapsto$ \edi/
                  ]
                  \}
            & \textit{// fixed assignments}
      \\
            & $\times$
                & \{
                  [
                  \minttok{\%0} $\mapsto$ \minttok{*}\esi/,
                  \minttok{\%2} $\mapsto$ \minttok{*}\esi/
                  ],
                  [
                  \minttok{\%0} $\mapsto$ \minttok{*}\ebp/,
                  \minttok{\%2} $\mapsto$ \minttok{*}\ebp/
                  ],
            & \multirow{2}{*}{\textit{// possible variations}} \\
            &   & \phantom{\{}
                  [
                  \minttok{\%0} $\mapsto$ \minttok{*}\esi/,
                  \minttok{\%2} $\mapsto$ \minttok{*}\ebp/
                  ],
                  [
                  \minttok{\%0} $\mapsto$ \minttok{*}\ebx/,
                  \minttok{\%2} $\mapsto$ \minttok{*}\ebx/
                  ], ...
                  \} \\
      \fsc/ & = & $\emptyset$ \\
      \fmbarrier/
            & = & \cinl{false}
    \end{tabular}
    \caption{Formal interface \fI/}
    \label{fig:formal_interface}
\end{figure}

\subsection{Interface compliance}
\label{sec:compliance}

An extended assembly chunk \fasme/ $\triangleq$ (\fcodet/, \fI/) is said to be {\it interface compliant}
if it respects both the {\it framing}  and the {\it unicity} conditions that we define below.

\paragraph{Observational equivalences}
As a first step, we  define  an equivalence relation \fequivtm/ over
memory states modulo a token assignment \fT/,
a set of observed tokens \fname{B}
and a memory separation flag \fname{F}.
We start by defining an equivalence relation \fequivt/.
We say that   \fmachine/$_1$ \fequivt/   \fmachine/$_2$ if,
for all \ftoken/ \ft/ in \fname{B},
\linebreak
\feval/(\fmachine/$_1$, \fget{\fT/}{\ft/}) = \feval/(\fmachine/$_2$, \fget{\fT/}{\ft/}).
We can generalize it to any pair of token assignments \fT/$_1$ and \fT/$_2$:
\fmachine/$_1$ \fequivtt/  \fmachine/$_2$ if,
for all \ftoken/s \ft/ in \fname{B}, \feval/(\fmachine/$_1$, \fget{\fT/$_1$}{\ft/}) = \feval/(\fmachine/$_2$, \fget{\fT/$_2$}{\ft/}).
Then, we  define an equivalence relation \fequivm/ over memory states.
We say that \fmachine/$_1$ \fequivm/ \fmachine/$_2$ if
for all (address) \flocation/ \fl/ in \faddress/, \fget{\fmachine/$_1$}{\fl/} = \fget{\fmachine/$_2$}{\fl/}.
The equivalence relation \fequivtm/ over memory states modulo a token assignment
\fT/ (which can be generalized to a pair \fT/$_1$ and \fT/$_2$ as above), a set
of tokens \fname{B} and a memory separation flag \fname{F} is finally defined as:

\noindent
\fmachine/$_1$ \fequivtm/ \fmachine/$_2$ if:
\fmachine/$_1$ \fequivt/ \fmachine/$_2$ $\land$ ($F$ = \cinl{false} implies
\fmachine/$_1$ \fequivm/ \fmachine/$_2$)

\paragraph{Framing condition}
The framing condition restricts what can be read and written by the assembly template.
Given a token assignment  \fT/,      we define a {\it location input} (resp.~{\it location output}) as a location pointed by a input (resp.~output) token.
Then the framing condition stipulates that:
\begin{enumerate*}
\item [(frame-read)] only initial values from input location can be read;
\item [(frame-write)] only clobber registers and location output are allowed to
  be modified by the assembly template.
\end{enumerate*}

More formally, a location is {\it assignable} if it can be modified
  (i.e.,  if it is mapped to by an output token \ft/, belongs to the clobber set
  \fsc/ or is a memory location \faddress/ when there is no separation
  $\neg\fmbarrier/$)%
, and {\it non-assignable}  otherwise. We then have:
\begin{description}[wide]
\item[frame-write] for all \fmachine/, for all \fT/ in \fst/,
for all non assignable \flocation/ \fl/:
    \fget{\fmachine/}{\fl/} = \fget{\fexec/(\fmachine/, \fsubst{\fcodet/}{\fT/})}{\fl/}.

\item[frame-read] for all \fmachine/$_1$, \fmachine/$_2$ and \fT/ in \fst/ such that \linebreak
\fmachine/$_1$ \fequivI/ \fmachine/$_2$:
\fexec/(\fmachine/$_1$, \fsubst{\fcodet/}{\fT/}) \fequivO/
 \fexec/(\fmachine/$_2$, \fsubst{\fcodet/}{\fT/}),
\end{description}

\paragraph{Unicity} 
Informally, the unicity condition is respected
when the evaluation of output tokens is independent from the chosen
token assignment.
More formally,
for all \fmachine/$_1$, \fmachine/$_2$ , \fT/$_1$ and \fT/$_2$ in \fst/ such that
\fmachine/$_1$ \fequivIT/ \fmachine/$_2$:

\begin{center}
  \fexec/(\fmachine/$_1$, \fsubst{\fcodet/}{\fT/$_1$}) \fequivOT/
  \fexec/(\fmachine/$_2$, \fsubst{\fcodet/}{\fT/$_2$}).
\end{center}
\noindent
Note that    \textbf{frame-read} is a sub-case of
unicity where \fT/$_1$ = \fT/$_2$.

\section{Check, patch and refine}
\label{sec:rustina}

\begin{figure*}[htbp]

\tikzset{
  -|-/.style={
    to path={
      (\tikztostart) -| ($(\tikztostart)!#1!(\tikztotarget)$) |- (\tikztotarget)
      \tikztonodes
    }
  },
  -|-/.default=0.5,
  |-|/.style={
    to path={
      (\tikztostart) |- ($(\tikztostart)!#1!(\tikztotarget)$) -| (\tikztotarget)
      \tikztonodes
    }
  },
  |-|/.default=0.5,
}

  \begin{tikzpicture}[
      arr/.style={->, black, shorten >= 1pt, shorten <= 1pt},
      box/.style={rectangle, draw, solid, rounded corners=1mm,minimum size=.7cm,
      inner sep=2pt},
      dec/.style={diamond,aspect=2.5,draw,solid,inner sep=2pt},
      tr/.style={font=\scshape\smaller[3]},
      title/.style={font=\large\scshape,fill=bggrey}
      ]

      \node[minimum height=2em] (asm) {GNU assembly template};
      \node[minimum height=2em,anchor=north] (fi) at (asm.south) {GNU assembly interface};
      \draw (asm.south west) -- (asm.south east);

      \node[fit=(asm)(fi), draw,
      inner xsep=2pt, inner ysep=1pt] (xasm) {};

      \node[anchor=east] (plus) at ($(xasm.west) + (left:.2)$) {+};
      \node[anchor=east] (c) at ($(plus.west) + (left:.2)$) {\sf C};

      \node[black!50,very thin,fit=(c)(xasm), draw, inner xsep=10pt, inner ysep=4pt, rounded corners=1mm] (c+asm) {};

      \node[minimum height=2em] (dba) at ($(asm)+(6.3,0)$) {IR template \fcodet/};
      \node[minimum height=2em] (formal interface) at ($(fi)+(6.3,0)$) {Formal interface \fI/};

      \node[fit=(dba)(formal interface), draw, inner xsep=2pt, inner ysep=1pt] (extracted) {};
      \draw (formal interface.north west) -- (formal interface.north east);

      \path (asm.east -| xasm.east) edge[arr] node[tr,above,pos=0.6] {Code semantics} node[tr,below,pos=0.6] {extraction}  (dba.west -| extracted.west);
      \path (fi.east -| xasm.east) edge[arr] node[tr,above,pos=0.6] {Interface semantics} node[tr,below,pos=0.6] {extraction}  (formal interface -| extracted.west);

      \node[dec,draw] (docheck) at ($(dba.north)+(3,0)$) {\scshape\small Check};
      \node[dec,draw] (dopatch) at ($(formal interface.south)+(3,0.25)$) {\scshape\small Patch};

      \path (dba -| extracted.east) edge[arr,-|-] (docheck.west);
      \path (formal interface -| extracted.east) edge[arr,-|-] (docheck.west);

      \node[align=left,anchor=west] (docheckresult) at ($(docheck)+(2,0)$) {Formally verified\\ compliance};
      \node[align=left,anchor=west] (dopatchresult) at ($(dopatch)+(2,0)$) {Formally compliant\\patch};
      \node[align=left,anchor=west] (nopatch) at ($(dopatch)+(2,-0.8)$) {Compliance alarm};

      \path  (docheck) edge[arr] node[auto] {\textcolor{darkred}{\faTimesCircle}} (dopatch);

      \path (docheck) edge[arr] node[near start,auto] {\textcolor{darkgreen}{\faCheckSquare}} (docheckresult);
      \path (dopatch) edge[arr] node[near start,auto] {\textcolor{darkgreen}{\faCheckSquare}} (dopatchresult);
      \draw[arr] (dopatch) |- node[near start,right] {\textcolor{darkred}{\faTimesCircle}} (nopatch);

    \end{tikzpicture}
    \vspace{-6mm}
  \caption{Overview of \proto/}
  \label{fig:flow_overview}
\end{figure*}

Figure~\ref{fig:flow_overview} presents an overview of \proto/. The
tool takes as input a C file containing inline assembly
templates in GNU syntax. From there, it parses the template code to
produce an {\it intermediate representation} (IR) of the template \fcodet/, and
interprets the concrete interface to produce a \textit{formal interface}
\fI/.
The tool then checks that the code complies with its interface
using dedicated {\it static dataflow analysis}.
If
it succeeds, we have {\it formally verified} that the assembly template
complies with its interface.
If not, our tool examines the difference between the formal interface
computed from the code and  the one extracted from specification;
it can then  produce a \textit{patch} (if some elements of the
interface were forgotten) or  \textit{refine} the interface (if the declared
interface is larger than needed). We cannot
produce a patch in every situation, in that case the tool reports
a compliance alarm --
they can be  false alarms, but it rarely happens on real
code.
Algorithms are fully detailed in the companion report~\cite{rustina_in_a_nutshell_2020}.

\subsection{Preliminary: code semantics extraction}
\label{sec:extraction}

\newcommand{\te}[1]{{\sf{#1}}}
\begin{figure}[htbp]
  \centering\scriptsize
  \begin{tabular}{rcl}
    \te{inst} & {\sf :=} & \te{lv} $\bm\leftarrow$ \te{e} | {\sf goto e}
                          | {\sf if} \te{e} {\sf then goto e}  {\sf else goto} \te{e} \\
    \te{lv} & {\sf :=} & {\sf var} | {\sf @[}\te{e}{\sf ]}$_{n}$ \\
    \te{e} & {\sf :=} & {\sf cst} | \te{lv}
                      | \te{unop} \te{e} | \te{binop} \te{e} \te{e}
                      | \te{e} {\sf ?} \te{e} {\sf :} \te{e} \\
    \te{unop} & {\sf :=} & ${\bm \neg}$ | ${\bm -}$ | {\sf zext$_{n}$}
                          | {\sf sext$_{n}$}    | {\sf extract$_{i .. j}$} \\
    \te{binop} & {\sf :=} & \te{arith} | \te{bitwise} | \te{cmp}
                           | {\sf concat} \\
    \te{arith} & {\sf :=} & ${\bm +}$ | ${\bm -}$ | ${\bm \times}$
                           | {\sf udiv}    | {\sf urem}
                           | {\sf sdiv}   | {\sf srem} \\
    \te{bitwise} & {\sf :=} & ${\bm \land}$ | ${\bm \lor}$ | ${\bm \oplus}$
                             | {\sf shl} | {\sf shr} | {\sf sar} \\
    \te{cmp} & {\sf :=} & ${\bm =}$ | ${\bm \neq}$
                         | $>_{u}$   | $<_{u}$
                         | $>_{s}$         | $<_{s}$
  \end{tabular}
  \caption{The \binsec/ intermediate representation}
  \label{fig:dba_ir}
\end{figure}

Our analyses rely on Intermediate Representations (IR) for binary code,
coming from  lifters~\cite{Brumley2011,DBLP:conf/cav/BardinHLLTV11} developed for
the context of binary-level program analysis. We use the IR
of the \binsec/ platform \cite{Djoudi2015,DBLP:conf/wcre/DavidBTMFPM16} (\cref{fig:dba_ir}), but
all such IRs are  similar. They  encode every machine code instruction into a small well-defined side-effect free language, typically
an imperative language over bitvector variables (registers) and arrays (memory),  providing jumps and conditional branches.
Still, code lifters do not operate directly on  assembly templates but on machine code,
requiring a little extra-work to recover the tokens.
We replace each token in the assembly template
by a distinct register,   use an existing assembler (\textsc{gas}) to
transform the new assembly chunk into machine code  and then lift it to
IR. We perform the whole operation again where each token
is mapped to another register, so as to distinguish tokens from
hard-coded registers.
%
Tokens are then replaced in IR by distinct new variable names.

\subsection{Compliance Checking}
\label{sec:check}

This section discusses our  static interface compliance checks.
We rely on the {\it dataflow analysis framework} \cite{10.1145/512927.512945}, intensively used in compilers and software verification.
We collect {\it sets of locations} (\ftoken/, \fregister/
or the whole \mintval{memory}) as dataflow facts, then compare them against the sets expected from the interface. Checking {\bf frame-write} requires a {\it forward impact analysis},
checking {\bf frame-read}  requires a {\it backward liveness analysis}, and finally {\bf unicity} requires a combination of both.
Our techniques are {\it over-approximated} in order to ensure soundness.
Memory is considered {\it as a whole} -- all  memory accesses being squashed as 
\mintval{memory}, with a number of advantages: it  closely follows the interface mechanisms for memory, helps termination (the set of dataflow facts is finite)
and saves us the complications of memory-aware static analysis (heap or points-to).
Finally, we propose two {\it precision optimizations}  in order to reduce the risk of false positives
(their impact is evaluated in \cref{sec:xps-cmp}).

\paragraph{Frame-write}
Check must ensure that non-assignable locations have the exact same value
before and after the execution.
As first approximation, a location that is never written (i.e., never on the Left Hand Side LHS
of an assignment) safely keeps its initial value -- since IR expressions are
side-effect free.
{\it Impact analysis} iterates forward from the entry of the chunk,
collecting the set of LHS locations
(either a \ftoken/, a \fregister/
or the whole \mintval{memory}
).
We then check that each LHS location belongs to the set of declared assignable
locations (i.e. \fbo/ $\cup$ \fsc/ together with
\mintval{memory} if $\neg\fmbarrier/$).

\paragraph{Frame-read}
Check must ensure that no uninitialized location is read.
This requires to compute (an overapproximation of) the set of \textit{live}
locations (i.e. holding a value that may be read before its next definition).
Liveness analysis   iterates backward from
the exit of the chunk, where output locations are live (outputs tokens \fbo/),
computing dependencies of the Right Hand Side (RHS) expression of
found definitions until the fix-point is reached.
We then check at the entry point that each live location
belongs to the set of declared inputs (i.e. \fbi/ together
with \mintval{memory} if $\neg\fmbarrier/$).

\paragraph{Unicity}
Check must ensure that compiler choices have no impact on
the chunk output.
What may  happen is that a location
is impacted or not  by a preceding write depending on the token assignment.
To check that this does not happen, we first define a relation
$\fsmash/$ over  $\flocationt/$ (incl.~tokens)
  such that
$\fl/\ \fsmash/\ \fl/'$ is false if we can prove that (writing on)  $\fl/$ has no
impact on (the evaluation of) $\fl/'$ -- whatever the token assignment.  In our implementation, $\fl/\ \fsmash/\ \fl/'$ returns
false if there is no token assignment where $\fl/$ is a
sub-expression of $\fl/'$.
Then, using previous \textbf{frame-write} and \textbf{frame-read} analyses,
we finally check at each assignment to a
location \fl/ that, for each live location \fl/', \fl/ \fsmash/ \fl/' returns
\cinl{false}.

We now sketch the implementation of  \fsmash/. The main challenge is to
avoid enumerating all valid token assignments \fst/ (c.f. \cref{sec:translate})
.
We compute over a smaller set of  abstract location facts \flocations/, indicating only
whether a location is a constant value
(\texttt{Immediate}), a register (\texttt{Direct} \fregister/) or is used to
compute the address of a token
(\texttt{Indirect} \fregister/).
We   {\it abstract token assignments} by reinterpreting the constraints over \flocations/,
yielding $\mathds{D}^*$ : \flocationt/ $\mapsto$ \flocations/ \cinl{set}.
We then  define  the relation $l^*$ \fsmashs/ ${l^*}'$ over \flocations/ as:

{
  \smaller
  \[
    l^* \text{ \fsmashs/ } {l^*}' =
    \begin{cases}
      \text{\texttt{Direct} } r \text{ \fsmashs/ \texttt{Direct} } r
      & : \cinl{true} \\
      \text{\texttt{Direct} } r \text{ \fsmashs/ \texttt{Indirect} } r
      & : \cinl{true} \\
      others & : \cinl{false}
    \end{cases}
  \]
}

Finally, we   build the relation \fl/ \fsmash/ \fl/' such that
it returns \cinl{true} (sound) except if
one of the following holds:
\begin{itemize}
\item no $l^*$, ${l^*}'$   in
  $\mathds{D}$(\fl/) $\times$ $\mathds{D}$(\fl/')
  such that $l^*$ \fsmashs/ ${l^*}'$;
\item \fl/ or \fl/' belongs to \fsc/;
\item \fl/ and \fl/' are tokens, \fl/ is early clobber (\cinl{"\&"});
\item \fl/ is equal to \fl/' (independent of compiler choice).
\end{itemize}

\begin{resbox}
  Our checkers are {\it semantically sound} in the sense that they compute an
  {\it overapproximation} of the assembly template semantics.
  Hence,  successfully checking an extended assembly chunk \textit{ensures}  it
  is  {\it interface-compliant}.
\end{resbox}

On the other hand, our technique could report compliance
issues that do not exist (false positives).
We propose below two {\bf precision improvements}:

\begin{description}[wide]
\item [1. Expression propagation]
  In  \cref{fig:motiv_c_src}, \textbf{frame-write} check, as is,
  would report a
  violation for \ebx/ and \esi/ because they are written.
  Yet, it is a false positive since both end up with their initial value.
  To avoid it, we perform a symbolic expression propagation for each
  written location,
  \textit{inlining} the definition of written locations into their RHS
  expressions,  and  performing {\it IR-level syntactic simplifications}
  -- such as $1+x-1 = x$ or $ x \oplus x = 0$.
  Then, at fixpoint, \textbf{frame-write} checks before raising an alarm whether
  the original value has been restored (no alarm) or not (alarm);

\item [2. Bit-level liveness dependency]
  In \cref{fig:motiv_c_src}, \cinl{result} takes only the lowest
  byte of \eax/. However, our basic technique
  will count both z and \eax/ as live while high bytes of \eax/ are
  actually not --  such imprecisions may  
lead  to false alarms (\cref{sec:xps-cmp}).
  We improve our liveness analysis to independently track
  the status of each location bit. For efficiency, we do not  propagate location bits
  but locations equipped with a bitset representing the status of each of their bits.
  We
  modify propagation rules accordingly (especially
  bit manipulations like extraction or concatenation), with bitwise operations over the bitsets.
\end{description}

\subsection{Interface Patching}
\label{sec:patch}

When the compliance checking fails,
\proto/ tries to generate a patch to fix the issue.
As our dataflow analysis infers an over-approximated interface
for the chunk under analysis,
we take advantage of it to strengthen the current interface.

\paragraph{Framing condition}
We build a patch that makes the
template \fcodet/ compliant with its formal interface \fI/ as follows:
\begin{description}[wide]
\item [frame-write] Any hard-coded register (resp. token) written without belonging to
  \fsc/ (resp. \fbo/) is added;

\item [frame-read]
  Any token read without belonging to \fbi/ and without being
  initialized before, is added
  .
  Reading a register before assigning it prevents automatic
  patch generation%
  \footnote{If this is done on purpose, the chunk actually is out of this paper's scope.}.
\end{description}
In both cases, any direct access to a memory cell sets memory separation \fmbarrier/
to \cinl{false}.

We then retrofit the changes of the formal interface in the concrete syntax
to produce the patch. For instance, in \cref{fig:motiv_patch},  token
\cinl{\%3} (i.e. \edx/) violates the \textbf{frame-write} condition.
We add a new output token \cinl{\%2} : \cinl{"=d" (dummy)} bound to its old
initializer : \cinl{"2" (old_val2)}. Since we add a new token, we take care to
keep template ``numbering'' consistent.

\begin{resbox}
  When a framing issue patch  is generated, the resulting chunk is {\it ensured} to meet the
  framing condition.
\end{resbox}

\paragraph{Unicity}
We give to the faulty register (resp.~token) the (resp.~early) clobber status
preventing it to be mis-assigned to another token.
Note however that, since we over-constrain the interface
(the syntax does not allow to declare a pair of entries as distinct),
the patch may fail if there is no more valid token assignment.

\begin{resbox}
  When a unicity patch is  generated,  the resulting chunk is {\it
    ensured} to be fully interface compliant if it
  still compiles.
\end{resbox}

\subsection{Bonus: Refining the interface}
\label{sec:refine}

 Even if overapproximated, the interface that is inferred by \proto/  during the
 check may be smaller than the declared one, allowing
 to produce  \textit{refinement patches} removing  unnecessary
 constraints in the interface -- which can in turn
 give more room to the compiler to produce smaller or faster
 code.

 We can already remove never-read inputs,
 never-written clobbers or undue \cinl{"memory"} keywords in absence of
 memory accesses%
 \footnote{These refinements can be disabled for dummy constraints put on purpose.}.
 There is another case where a \cinl{"memory"} constraint can be removed.
 Indeed, as recommended in the documentation, single-level pointer accesses
 can be declared by common entries using the \cinl{"m"} placement constraint
 instead of the (much more expensive) \cinl{"memory"} keyword.

 We design a dedicated ``points-to'' analysis to identify the candidates
 for this transformation. It is based on a dataflow analysis collecting,
 for each memory access, the precise location (on the form
 {\it token or symbol + offset}) and size of the access.
 If it succeeds, we can safely remove the \cinl{"memory"} keyword and
 instead add a new entry (input \cinl{"m"}, output \cinl{"=m"} or both
 depending of the access pattern) for each of the identified base pointers.

\begin{figure}[htbp]
   \centering
  \includegraphics[trim=0 0 120 0, clip]{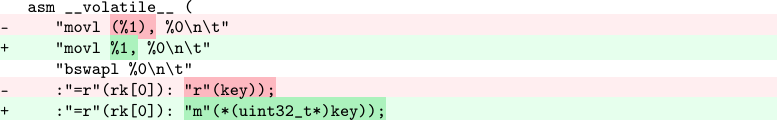}

\caption{Smart patch of a \libtomcrypt/ chunk}
  \label{fig:refine_memory}
\end{figure}

 \Cref{fig:refine_memory} shows an example of refinement happening in
 \libtomcrypt/. In the original code, the \cinl{"memory"} constraint was
 forgotten. We can see that (patch) refinement produces a fix that
 does not add the missing keyword, but instead changes
 the way the content pointed by \cinl{key} is given to the chunk.

\section{Implementation}
\label{sec:implementation}

We have implemented \proto/, a prototype for interface compliance analysis
following the method described in \cref{sec:rustina}.
\proto/ is written in \caml/ ($\sim$3 kLOC), it  is
based on
\framac/~\cite{DBLP:journals/fac/KirchnerKPSY15} for C manipulation
(parsing, localization and patch generation),
\binsec/ \cite{Djoudi2015,DBLP:conf/wcre/DavidBTMFPM16} for IR lifting (including basic syntactic simplifications),
and \textsc{GAS} to translate assembly into machine code.
Our tool can handle a large portion of the \x86/ and \arm/
instruction sets.
Yet, float and system instructions are not supported (they are unsupported
by \binsec/). Despite this, we
handle 84\% of assembly chunks found in a Debian distribution
(\cref{sec:evaluation}).

\section{Experimental evaluation}
\label{sec:evaluation}

\paragraph{Research questions}  We consider 5  research questions%
\begin{enumerate*}[label={\bf RQ\arabic*.},ref={\bf RQ\arabic*}, series=rqitems]

\item \label{item:rq1} Can \proto/ automatically check interface compliance on  assembly chunks found in the wild?

\item \label{item:rq2} Incidentally, how  many assembly chunks exhibit a
  compliance issue, and which ones are the most frequent?

\item \label{item:rq3} Can \proto/ automatically patch detected compliance issues?

\item \label{item:rq4}  What is the real impact of the compliance issues reported and of the generated patches?

\item \label{item:rq5}  What is the impact of \proto/ design choices on the overall checking result?

\end{enumerate*}

\label{sec:xps}

\paragraph{Setup}
All experiments are run on a regular \textsf{Dell Precision 5510} laptop
equipped with an \textsf{Intel Xeon E3-1505M} v5 processor and
32GB of RAM.

\paragraph{Benchmark} We run our prototype on
\textit{all} C-related \textbf{\x86/} inline assembly chunks  found in a  {\it Linux Debian 8.11 distribution},
i.e.,  3107  \textbf{\x86/} chunks in 202 packages,
 including big inline assembly users like
\alsa/, \gmp/ or \ffmpeg/.
We remove 451 out-of-scope chunks (i.e., containing either float or system
instructions),  keeping {\it 2656 chunks} (85\% of the initial dataset),
with mean
size of 8 assembly instructions (max.~size: 341).

\begin{table}[htbp]
  \caption{\proto/ application on Debian 8.11 \x86/}
  \label{tbl:patch_diff}

  \begin{subtable}[b]{.5\textwidth}
    \caption{Overview at package level}
    \label{tbl:patch_diff_package}

    \pgfplotstableread[row sep=\\,col sep=&]{
      kind  & compliant & benign & serious  \\
      0     & 58      & 15     & 27      \\
      1     & 88      & 0      & 12       \\
    }\patchdiff
    \centering
    \begin{tabular}{lrrcrrc}
      \multirow{2}{*}{\textbf{Packages} considered}
      & \multirow{2}{*}{\textbf{202}}
      & \multicolumn{3}{l}{\smaller[2] average chunks}
      & \smaller[2] 15 \\
      &
      & \multicolumn{3}{l}{\smaller[2] max chunks}
      & \smaller[2] 384
      \\
      \cmidrule(lr){1-7}
      \addlinespace
      & \multicolumn{2}{c}{Initial} & \tikzmark{dap}
      & \multicolumn{2}{c}{Patched} &
      \\
      \cmidrule(lr){2-3}
      \cmidrule(lr){5-6}
      \enspace \makebox[\widthof{\faQuestionCircle}]
      {\textcolor{darkgreen}{\faCheckSquare}} -- fully compliant
      & 117 & \smaller 58\% & & 178 & \smaller 88\% &
      \\
      \enspace \makebox[\widthof{\faQuestionCircle}]
      {\textcolor{orange}{\faShield}} -- only benign issues
      & 31 & \smaller 15\% & & 0 & \smaller 0\% &     \\
      \enspace \textcolor{darkred}{\faTimesCircle} -- serious issues
      & 54 & \smaller 27\% & & 24 & \smaller 12\% &
    \end{tabular}

    \begin{tikzpicture}[overlay, remember picture]
      \coordinate (x) at (pic cs:dap);
      \coordinate (y) at ($(x) + (left:.2) + (down:1.12)$);
      \begin{axis}[
        at=(y),
        width=115pt,
        height=70pt,
        axis y line=none,
        axis x line=none,
        bar width=5pt,
        ybar stacked,
        ]
        \addplot[fill=darkred]
        table[x=kind,y=serious]{\patchdiff};
        \addplot[fill=orange]
        table[x=kind,y=benign]{\patchdiff};
        \addplot[fill=darkgreen]
        table[x=kind,y=compliant]{\patchdiff};
      \end{axis}
    \end{tikzpicture}

  \end{subtable}

  \begin{subtable}[b]{.5\textwidth}
    \bigskip
    \caption{Overview at chunk level}
    \label{tbl:patch_diff_chunk}

    \pgfplotstableread[row sep=\\,col sep=&]{
      kind  & compliant & benign & serious  \\
      0     & 49      & 40   & 11      \\
      1     & 97      & 0    & 3       \\
    }\patchdiff

    \centering
    \begin{tabular}{lrrcrrc}
      Assembly chunks
      & 3107
      \\
      \quad \smaller out-of-scope (e.g.~floats)
      & 451
      \\
      \multirow{2}{*}{\textbf{Relevant} chunks}
      & \multirow{2}{*}{\textbf{2656}}
      & \multicolumn{3}{l}{\smaller[2] average size}
      & \smaller[2] 8
      \\
      &
      & \multicolumn{3}{l}{\smaller[2] max size}
      & \smaller[2] 341
      \\
      \cmidrule(lr){1-7}
      \addlinespace
      & \multicolumn{2}{c}{Initial} & \tikzmark{dac}
      & \multicolumn{2}{c}{Patched} &
      \\
      \cmidrule(lr){2-3}
      \cmidrule(lr){5-6}
      \enspace \makebox[\widthof{\faQuestionCircle}]
      {\textcolor{darkgreen}{\faCheckSquare}} -- fully compliant
      & 1292 & \smaller 49\% & & 2568 & \smaller 97\% &
      \\
      \enspace \makebox[\widthof{\faQuestionCircle}]
      {\textcolor{orange}{\faShield}} -- only benign issues
      & 1070 & \smaller 40\% & & 0 & \smaller 0\% &     \\
      \enspace \textcolor{darkred}{\faTimesCircle} -- serious issues
      & 294 & \smaller 11\% & & 88 & \smaller 3\% &
    \end{tabular}

    \begin{tikzpicture}[overlay, remember picture]
      \coordinate (x) at (pic cs:dac);
      \coordinate (y) at ($(x) + (left:.2) + (down:1.17)$);
      \begin{axis}[
        at=(y),
        width=115pt,
        height=71pt,
        axis y line=none,
        axis x line=none,
        bar width=5pt,
        ybar stacked,
        ]
        \addplot[fill=darkred]
        table[x=kind,y=serious]{\patchdiff};
        \addplot[fill=orange]
        table[x=kind,y=benign]{\patchdiff};
        \addplot[fill=darkgreen]
        table[x=kind,y=compliant]{\patchdiff};
      \end{axis}
    \end{tikzpicture}

  \end{subtable}

  \begin{subtable}[b]{.5\textwidth}
    \bigskip
    \caption{Overview of found issues}
    \label{tbl:patch_diff_issue}

    \centering
    \begin{tabular}{lrrrr}
      & \multicolumn{2}{c}{Initial}
      & \multicolumn{2}{c}{Patched}
      \\
      \cmidrule(lr){2-3}
      \cmidrule(lr){4-5}
      \textbf{Found} issues
      & \textbf{2183} &
      & 183 &
      \\
      \quad significant issues
      & 986 &
      & 183 &
      \\
      \addlinespace
      \textbf{frame-write}
      & \textbf{1718} &
      & 0
      \\
      \enspace \makebox[\widthof{\faQuestionCircle}]
      {\textcolor{orange}{\faShield}} -- flag register clobbered
      & 1197 & \smaller 55\% & 0 & \smaller 0\%
      \\
      \enspace \textcolor{darkred}{\faTimesCircle} --
      read-only input clobbered
      & 17 & \smaller 1\% & 0 & \smaller 0\%
      \\
      \enspace \textcolor{darkred}{\faTimesCircle} --
      unbound register clobbered
      & 436 & \smaller 20\% & 0 & \smaller 0\%
      \\
      \enspace \textcolor{darkred}{\faTimesCircle} --
      unbound memory access
      & 68 & \smaller 3\% & 0 & \smaller 0\%
      \\
      \addlinespace
      \textbf{frame-read}
      & \textbf{379} &
      & 183
      \\
      \enspace \textcolor{darkred}{\faTimesCircle} --
      non written write-only output
      & 19 & \smaller 1\% & 0 & \smaller 0\%
      \\
      \enspace \textcolor{darkred}{\faTimesCircle} --
      unbound register read
      & 183 & \smaller 8\% & 183 & \smaller 100\%
      \\
      \enspace \textcolor{darkred}{\faTimesCircle} --
      unbound memory access
      & 177 & \smaller 8\% & 0 & \smaller 0\%
      \\
      \addlinespace
      \textbf{unicity}
      & \textbf{86} & \smaller 4\%
      & 0 & \smaller 0\%
      \\
    \end{tabular}

  \end{subtable}

\end{table}

\subsection{Checking (\rqcheck/,\rqbugs/)} \label{sec:xp:check}

\cref{tbl:patch_diff} sums up compliance checking results
before (``Initial'') and after patching (``Patched'') -- we focus here on the Initial case.

\paragraph{Results}
\proto/ reports in less than 2 min (\textit{40 ms per chunk in average}) that
1292 chunks out of 2656 are (fully) interface compliant (resp.~117 packages out of 202), while 1364 chunks (resp.~85 packages) have compliance issues.
Among the noncompliant ones, \proto/ allows to pinpoint 294 chunks (resp.~54 packages) with  serious compliance issues -- according to our study in \cref{sec:assessment}
we count an issue as benign  only when it corresponds to missing the flag register as clobber (P1 in \cref{sec:assessment}).

\paragraph{Quality assessment}
While chunks deemed compliant by \proto/ are indeed supposed to be compliant (yet, it is still useful to test it),
compliance issues could be false alarms.

We evaluate these two cases with 4 elements.
\begin{enumerate*}[label=(\(\bf qa_{\arabic*}\)), ref={\(\bf qa_{\arabic*}\)}]
\item \label{criteria:replay}
  We run \proto/ on known \libatomic/ and \glibc/  compliance bugs and on their patched versions
  : every time, \proto/ returns the expected
  result.
\item \label{criteria:projects}
  We consider 8 significant projects (\cref{sec:real-life-impact}), {\it
  manually review all their faulty assembly chunks} (covering roughly 50\% of
 the serious issues reported in \cref{tbl:patch_diff_issue}) as well as  randomly chosen compliant chunks,  and crosscheck results
with \proto/: they perfectly match.
\item \label{criteria:postcheck} For compliance proofs, we also run the checker
  after patching: \proto/ deems all patched chunks compliant.
\item \label{criteria:realfixes} Several patches sent to developers have been
  accepted (\cref{sec:xp:impact}).
\end{enumerate*}

\textit{We conclude that results returned by \proto/ are good: as expected, a chunk deemed compliant is compliant, and reported compliance issues
are most likely true alarms -- we do not find any false alarm in our dataset.}

\paragraph{\arm/ benchmark}
We also run \proto/ on the \arm/ versions of \ffmpeg/,
\gmp/ and \libyuv/ (from \textit{Linux Debian 8.11})
for a total of 394 chunks (average size 5, max.~size 29).
We found very few issues (78), all in   \ffmpeg/ and  related to the use
of  special flag \mintflg{q} (accumulated saturations).
Manual review confirms them.
Interestingly, the \cinl{"cc"} keywords are not forgotten in other cases.
As flags are explicit in \arm/ mnemonics, coding practices are different
than those for \x86/.

\begin{resbox}
\rqcheck/: \proto/ is effective at compliance checking, in terms of speed
and  precision -- yielding  compliance proofs and  identifying compliance bugs with near-zero false alarm rate.
 \proto/ is widely applicable: it runs on the full Debian assembly chunk
 base and, without change, on 2 different architectures.
\end{resbox}

\paragraph{Compliance bugs in practice} Our previous precision analysis allows
to assume that
 a warning from the checker likely indicates a true compliance issue.
Hence, according to \cref{tbl:patch_diff_chunk,tbl:patch_diff_package},
1364/2656 chunks (resp.~85/202 packages) are not interface-compliant, and 294
chunks  (resp.~54 packages) have significant issues.
According to \cref{tbl:patch_diff_issue},
 53\% of significant issues come from
\textit{unexpected} writes,  38\% from
\textit{unexpected} reads while 9\% are unicity problems.

\begin{resbox}
\rqbugs/: About half of  inline \x86/ assembly chunks found in the wild is {\it not} interface-compliant,  and a significant part (11\%) even exhibits significant compliance issues -- affecting 27\% of the packages under analysis.
\end{resbox}

\begin{table*}[!htbp]
  \caption{Inline assembly recurrent (compliance) error patterns}
  \label{tbl:clobber-omission}

  \centering
  \begin{tabular}{rlllllll}
     \makebox[0pt]{Pattern\ \ \ \ \ }  &    Omitted clobber  & Additional context                &  Implicit protection
                                                                                       &
                                                                                         Details
                                                                                                                    &
                                                                                                                      Robust?
       &   \# issues  &  Known bug   \\[3pt]

    P1 & \cinl{"cc"}        & --      & compiler choice                   &  \cinl{"cc"} clobbered by default                      &  \yes/ (*)           &  1197   & -- \\
    P2 & \ebx/ register             & --      & compiler choice                   &  \ebx/ protected in PIC mode        &  \no/ (\gcc/ $\geq$ 5)       &  30      & \cite{libatomic_ops6a7ee55}  \\
    P3 & \esp/ register             & \cinl{push/pop}      & compiler choice                   &  \esp/ protected                    &  \no/ (\gcc/ $\geq$ 4.6)     &  5       & \cite{libatomic_ops04eeeb6} \\
    P4 & \cinl{"memory"}    & single-chunk function      & function embedding                &  functions treated separately       &  \no/ (inlining, cloning)    &  285     & \cite{libtomcrypt_cefff85} \\
    P5 & MMX register               & single-chunk function       & ABI                &  MMX are ABI caller-saved           &  \no/ (inlining, cloning)    &  363      &  -- \\
    P6 & XMM register               &   disable XMM     & compiler option                   &  no XMM generation             &  \no/ (cloning) &  109      &  --  \\
  \end{tabular}

\smallskip

(*) There are discussions on \gcc/ mailing list to change that~\cite{68095c4_gcc_bugzilla_2015}.

\end{table*}

\subsection{Patching  (\rqpatch/)} \label{sec:xp:patch}

\paragraph{Results}
 \cref{tbl:patch_diff} (column ``Patched'') shows that \proto/ \textit{performs well} at patching
compliance  issues:
in \textit{2 min}, it patches
{\bf 92\%} of total issues ({2000/2183}), including  {\bf 81\%} of significant
issues (803/986).
Overall, 1276 more chunks (61 more packages) become
fully compliant, reaching  {\bf 97\%}  compliance on chunks  (\textbf{88\%}
on packages).

The remaining issues (unbound register reads) are out of the scope of patching. They often correspond to the case where
some registers are used as global memory between assembly chunks while only C
variables can be declared as input in inline assembly. This practice is however
 fragile (special case of pattern P6 in \cref{sec:assessment}).

\paragraph{Quality assessment}  We assess the quality of the patches adapting
\ref{criteria:replay} and \ref{criteria:projects} from \cref{sec:xp:check} as follows:
\begin{enumerate*}[label=(\textbf{$\bf qa'_{\arabic*}$})]
\item On known \libatomic/ and \glibc/ compliance bugs,
  comparing \proto/-generated patches to originals shows that they are functionally
  equivalent, with similar fixes.
\item We {\it manually review all (114) generated patches} on 8
  significant projects (\cref{sec:real-life-impact}) and check that they do
  fix the reported compliance issues.
\end{enumerate*}
Also, recall that patched chunks pass the compliance
checks (\ref{criteria:postcheck}) and that several patches
have been accepted by
developers (\ref{criteria:realfixes}).
Overall, \textit{in most cases our automatic patches are optimal and
  equivalent to the ones that would be written by a human}.
Still, the \cinl{"memory"} keyword may have a significant impact on
performance and developers usually try to avoid it. 
We address this issue with refinement  (\cref{sec:refine}).
Finally, some  unicity issues we found were actually resolved by
developers by (deeply) rewriting the assembly template, instead of
simply patching the interface.

\begin{resbox}
\rqpatch/: \proto/ effectively generates patches for compliance issues, in
terms of speed and patch quality. \proto/ can automatically curate a large code base, removing the vast majority of compliance issues --   the remaining ones require  rewriting  the code beyond mere interface compliance.
\end{resbox}

\subsection{Real-life impact (\rqimpact/)} \label{sec:xp:impact}
\label{sec:real-life-impact}

We have selected \nbsubmittedsolvedprojects/ significant projects from our benchmark (namely: \alsa/, \ffmpeg/, \haproxy/, \libatomic/, \libtomcrypt/, \udpcast/, \xfstt/, \px264/) to submit patches generated by \proto/ in order to get  real-world feedback.
Note that  submitting patches is time-consuming:
patches must adhere to the project policy and
our generated patches cannot be directly applied when the code uses
macros (a common practice in inline assembly) as \proto/ works on
preprocessed C files.

\Cref{tbl:patches} presents our results.
Overall, we submitted {\bf \nbsubmittedpatchs/} patches fixing
{\bf \nbsubmittedsolvedissues/} issues in the
{\bf \nbsubmittedsolvedprojects/} projects. 
  Feedback has been very positive:
  {\bf \nbacceptedpatchs/} patches have already been
  integrated, fixing {\bf \nbacceptedsolvedissues/} issues in
  {\bf \nbacceptedsolvedprojects/} projects
  (\alsa/, \haproxy/, \libatomic/, \libtomcrypt/, \udpcast/, \xfstt/, \px264/)
  -- developers clearly expressed their interest in using \proto/ once released.  The \ffmpeg/ patches are still under review.

\begin{table}[tbp]
    \caption{Submitted patches}
    \label{tbl:patches}
    \smaller
    \hspace{-.5cm}
    \begin{threeparttable}
      \begin{tabular}{llcccc}
        & & & Patched  & Fixed & \\
            Project     & About      & Status       & chunks & issues & Commit \\
            \cmidrule(lr){1-1}\cmidrule(lr){2-2}\cmidrule(lr){3-3}\cmidrule(lr){4-4}\cmidrule(lr){5-5}\cmidrule(lr){6-6}

          \alsa/        & Multimedia      & Applied & 20 & 64/64 & \href{https://github.com/alsa-project/alsa-lib/commit/01d8a6e03a4c1055e5c0ef6d5b6cfdadce545007}{01d8a6e}, \href{https://github.com/alsa-project/alsa-lib/commit/0fd7f0cdc5e663e69486d17b0207434396620be6}{0fd7f0c} \\
          \haproxy/     & Network         & Applied & 1  & 1/1 & \href{https://github.com/haproxy/haproxy/commit/09568fd54d2f091860cafa5173893445cd55c44c}{09568fd} \\
          \libatomic/   & Multi-threading & Applied & 1  & 1/1 & \href{https://github.com/ivmai/libatomic_ops/commit/d728ce4e2be5c8328f0af8fc738622915c520aee}{05812c2} \\
          \libtomcrypt/ & Cryptography    & Applied & 2 & 2/2 & \href{https://github.com/libtom/libtomcrypt/commit/cefff85550786ec869b39c0cb4a5904e88c84319}{cefff85} \\
          \udpcast/     & Network         & Applied & 2  & 2/2 & \href{http://www.udpcast.linux.lu/changes.html}{20200328} \\
        \xfstt/       & X Server        & Applied & 1  & 3/3 & \href{https://github.com/guillemj/xfstt/commit/91c358eeb4380e8235c66fa15456a039ff869509}{91c358e} \\
        \px264/       & Multimedia      & Applied & 11 & 83/83 & \href{https://code.videolan.org/videolan/x264/-/merge_requests/36}{69771} \\
          \ffmpeg/      & Multimedia      & Review  & 76 & 382/382\tnote{1}  \\

        \end{tabular}
    \begin{tablenotes}
    \item[1] Including 27 non automatically patchable issues, manually fixed.
  \end{tablenotes}
\end{threeparttable}

\end{table}

\begin{resbox}
\rqimpact/: \proto/ helps  efficiently  deliver quality patches.
\end{resbox}

\subsection{Internal evaluation: 
  precision optimizations 
  (\rqinternal/)}
\label{sec:xps-cmp}

The observed absence of false positives in \cref{sec:xp:check} already takes
into account the two precision enhancers (bit-level liveness analysis and symbolic expression propagation)  presented
in \cref{sec:check}.
We seek now to assess the impact of these two improvements over the false positive rate (fpr) of \proto/.
We ran a basic version of \proto/  (no expression propagation, no bit-level liveness, but still the basic IR simplifications done by \binsec/) on our whole benchmark.
It turns out that this basic version reports 127 false alarms (6\% fpr)   -- 40 \textbf{frame-write} (2\% fpr) and  87  \textbf{frame-read} (23\% fpr).
All these alarms concern potentially significant issues.
Restricting to  significant issues, this amount to false positive rates of
13\% (total), 23\% (\textbf{frame-read}) and 8\% (\textbf{frame-write}).
It turns out that our  two optimizations are complementary: bit-level liveness analysis removes the 87   false \textbf{frame-read} alarms while expression propagation removes the 40 false
\textbf{frame-write} alarms.

\begin{resbox}
  The two precision optimizations (expression folding, bit-level liveness) upon \proto/ base technique  are essential
  in order to get a near-zero false alarm rate.
\end{resbox}

\section{
  Bad coding practices for inline assembly}
\label{sec:assessment}

In this section, we aim to:
\begin{enumerate*}
\item seek some sort of regularity
 behind so many compliance issues, in order to
understand while developers introduce them in the first place;
\item understand in the same time why so many compliance issues do not turn more often
  into observable bugs;
\item assess the risk of such  bugs to occur in the future.
\end{enumerate*}

\paragraph{Common error patterns for inline assembly}
We have identified 6 patterns (P1 to P6, see \cref{tbl:clobber-omission})
responsible for  91\%
of  compliance issues (1986/2183) --  80\% of significant compliance issues (789/986).
In each case, some input or output declarations are missing, but surprisingly it almost always concerns the same registers
(\ebx/, \esp/, \cinl{"cc"}, MMX or XMM registers) or  {\tt memory}, with similar coding practices (e.g.~no  XMM declaration together with compiler options for deactivating
XMM, or no declaration of \ebx/ together with surrounding \asminl{push} and \asminl{pop}).
Hence, these patterns are  deliberate rather than mere  coding errors.

\paragraph{Underlying implicit protections and their limits}
It turns out that each pattern builds on implicit protections (\cref{tbl:clobber-omission}).
We identified three main categories:
\begin{enumerate*}[label=(\textbf{\arabic*})]
\item (P1-P2-P3)  compiler choices regarding inline assembly (e.g., ``protected''
  registers, default clobbers);
\item (P4-P5) the apparent protection offered by putting a single assembly chunk inside
  a C function (relying mostly on the limited interprocedural analysis abilities
  of compilers); and
\item (P6) specific compiler options.
\end{enumerate*}

{\it Yet, all these reasons are fragile}:  compiler choices may change, and  actually do, compilers enjoy more and more powerful  program analysis
engines including very aggressive code inlining like Link-time optimization (LTO),
and refactoring may affect the compilation context. %

We now provide a precise analysis of each error pattern:

\begin{enumerate}[label={P\arabic*},ref={P\arabic*}]

\item \textbf{omitted \cinl{"cc"} keyword.}  \x86/ has been once a ``cc0'' architecture,
  i.e., any
  inline assembly statement implicitly clobbered \cinl{"cc"}  so it was not
  necessary to declare it as written. As far as we know, compilers still unofficially
  maintain this special treatment for backward compatibility.
  However, some claim
  \textquote{that is ancient technology and one day it will be gone
    completely, hopefully}~\cite{68095c4_gcc_bugzilla_2015};

\item \textbf{omitted \ebx/ register.}
  The Intel ABI states  that \ebx/ should be  treated
  separately as a special PIC (Position Independent Code) pointer register.
  Old version of \gcc/ (prior to version < 5.0) totally dedicated
  \ebx/ to that role and refrained from binding  it to an assembly chunk.
  Still, some chunks actually require to use \ebx/ (e.g. \asminl{cmpxchg8b})
  and people used tricks to use it anyway without stating it.
  It becomes risky because current compilers 
  can now spill and use \ebx/ as they need;

\item \textbf{omitted \esp/ register.}
  \esp/ is here modified but  restored   by \mintkey{push} and \mintkey{pop}.
  Yet, compilers may decide to use \esp/ instead of \ebp/ to pass addresses of
  local variables.
  In fact, it became the default behavior since \gcc/ version 4.6.
  Thus, code mixing local variable references and \mintkey{push} and
  \mintkey{pop} may read the wrong index of the stack, leading to
  unexpected issues;

\item \textbf{omitted \cinl{"memory"}.}
  Compilers' analysis
  are often performed {\it per} function, with conservative assumptions on the memory impact of called functions,
limiting the ability of the compiler to
  modify (optimize) the context of chunks.  This is no longer true in case of inlining  where
assembly interface issues become more impactful;

\item \textbf{omitted MMX register.}
  For the same reason as above, when a chunk is inside a function,
  it is also protected by the ABI in use. The Intel ABI specifies
  that MMX registers are caller-saved, hence the compiler must ensure that their
  value is restored when function exits.
  Yet, inlining may break this pattern since the ABI barrier is not there  anymore once
  the function code is inlined;

\item \textbf{omitted XMM register.}   Using parts of the architecture out-of-reach of the
compiler (the compiler cannot spill them, typically  through adequate command-line options)
is  safe but fragile as  it is sensitive to future refactoring (affecting the compiler options).
Moreover, newer compiler options or hardware architecture updates can implicitly reuse registers otherwise deactivated, e.g.~ XMM registers
  reused as subpart of  AVX registers.

\end{enumerate}

\paragraph{
  Breaking patterns}
We now seek to assess how fragile (or not) these patterns are.
Replaying known
issues~\cite{libatomic_ops6a7ee55,libatomic_ops04eeeb6,libtomcrypt_cefff85}
with current compilers shows that patterns P2 to P4 are (still) unsafe.
In addition, we conducted experiments to show that current compilers
do have the technical capacity to break the patterns.
We consider two main threat scenarios:
\begin{description}[leftmargin=2pt]

\item [Cloning]
  developers copy the chunk as is to another project
  (bad but common development practice~\cite{10.1145/1984701.1984706,10.1145/2901739.2901767});

\item [Inlining]
  projects import the code as a library and compile it statically with
  their code (link-time optimization).
\end{description}

We consider for each pattern 5 representative faulty chunks from the 8 projects.
For each chunk, we craft  a toy example  aggressively tuned to call  the (cloned or imported) chunk
in an optimization-prone context.
For instance, as P5 \& P6 issues involve SIMD registers, the corresponding chunks are called
within an inner loop while \textit{auto-vectorization} is enabled
(\texttt{-O3}).
Results are reported in column ``Robust?'' of \cref{tbl:clobber-omission}.
We actually break
  5/6 patterns with code cloning (all but P1),
 and 4/6 with code inlining, demonstrating that these compliance issues
should  be considered   plausible threats.

\begin{resbox}
We identified a set of 6 recurring patterns leading to the majority of compliance issues.
All of them  build on fra\-gile assumptions on the compiling chain.
Especially, code cloning and compiler code inlining are serious threats.
\end{resbox}

\section{Discussion}
\label{sec:discussion}

\subsection{Threats to validity}  \label{app:sec:threats-validity}  

We avoid bias as much
as possible in our benchmark:
\begin{enumerate*}
\item the benchmark is comprehensive:   all  Debian packages with C-embedded
  inline assembly;
\item we mostly work on  x86, but still consider 394 ARM chunks from 3 popular projects.
\end{enumerate*}
Our prototype is based on  tools already used in  significant case studies~\cite{bardin_backward-bounded_2017,david_specification_2016,DanielBR20,FeistMBDP16}, including a well tested x86-to-IR decoder \cite{DBLP-conf/kbse/KimFJJOLC17}.
Also, results have been crosschecked in several ways
and some of them manually reviewed.
So, we feel confident in our main conclusions.

\subsection{Limitations}

\paragraph{Architecture}
Our  implementation supports the architectures of  the \binsec/ platform, currently
\x86/-32 and \arm/v7.
This is not a conceptual limitation, as our technique ultimately works on a generic  IR.
As soon as a new architecture is available in \binsec/,  we will support it for
free.

\paragraph{Float} We do not yet support float instructions  as \binsec/ IR does not.
While adding support in the IR is feasible but time-consuming, our technique could also work solely with a
{\it partial instruction support} reduced to I/O information about each instruction -- at the price of some false positives.

\paragraph{System instructions}
Our formalization considers assembly chunks as a deterministic way to convert well-identified inputs from the C environment   to
outputs.
But system instructions often read or write locations hidden to the C context  (system registers) and  will thus  appear to be
non-deterministic -- breaking either the framing or the unicity condition.
Extending our formalization
is feasible, but it is useful only if the GNU syntax is updated.
Still, we consider that at most  13\% of assembly chunks used such instructions.

\subsection{Microsoft inline assembly} \label{sec:masm}
Microsoft inline assembly (inline MASM) proposed in \visualstudio/~\cite{vsdoc}   does not suffer
from the same flaws as \gnu/'s. Indeed, each assembly instruction is known by
the compiler such that {\it no interface is required},
and moreover developers can seamlessly write variables from C
into the assembly mnemonics.
Yet, this solution is actually restricted to a subset of the i386
instruction set, as
the cost in term of compiler development is significantly
more important.

\section{Related work}

\paragraph{Interface compliance}
Fehnker et al.~\cite{Fehnker08} tackle inline assembly compliance {\it checking}
for ARM  (patching and refinement are not addressed), but in a very
limited way.
This work restricts compliance to the framing case (no unicity
condition) and is driven by assembly syntax rather than semantics,
making it less precise than ours -- for example, a
saved-and-restored register will be counted as a framing-write issue.
Moreover, it  does not handle neither memory  nor token constraints
(tokens are assumed to be in registers and to be distinct
from each other).
Finally, their implementation is strongly tied to \arm/ with strong syntactic
assumptions
and their prototype is evaluated only on 12 files
from a single project.

\paragraph{Assembly code lifting and mixed code verification}
Two  recent  works \cite{8952223,DBLP:conf/uss/CorteggianiCF18} lift GNU inline assembly
to semantically equivalent  C code in order to perform verification of  mixed
codes combining C and inline assembly. Their work is  complementary to ours: their lifting  {\it assume} interface compliance but in turn
 they can prove functional correctness of assembly chunks.
Verifying code mixing C and assembly has also been active
on Microsoft MASM assembly~\cite{Maus2008,DBLP:phd/dnb/Maus11,Schmaltz2012}.
Yet, inline MASM does not rely on interface (\cref{sec:masm}).

\paragraph{Binary-level analysis}
While binary-level semantic analysis is hard
\cite{DBLP:journals/toplas/BalakrishnanR10,DBLP:conf/fm/DjoudiBG16,DBLP:conf/vmcai/BardinHV11,DBLP-conf/vmcai/KinderK12},  inline assembly
chunks offer nice structural properties~\cite{8952223}
allowing efficient and precise analysis.
We also benefit from previous  engineering efforts on
generic binary lifters
\cite{Brumley2011,DBLP:conf/cav/BardinHLLTV11,DBLP-conf/kbse/KimFJJOLC17}.

\section{Conclusion}

Embedding GNU-like inline assembly into higher-level languages such as C/\Cpp/
allows higher performance, but at the price of potential errors  due
either to the assembly glue or to undue code optimizations as the
compiler blindly trusts the assembly interface. We propose a novel
technique to automatically reason about inline assembly interface compliance,
based on a clean formalization of the problem.
The technique is implemented in  \proto/, the first sound tool providing
comprehensive automated interface compliance checking as well as automated
patch synthesis and interface refinements.

\clearpage

\bibliography{biblio}

\begin{thebibliography}{10}
\providecommand{\url}[1]{#1}
\csname url@samestyle\endcsname
\providecommand{\newblock}{\relax}
\providecommand{\bibinfo}[2]{#2}
\providecommand{\BIBentrySTDinterwordspacing}{\spaceskip=0pt\relax}
\providecommand{\BIBentryALTinterwordstretchfactor}{4}
\providecommand{\BIBentryALTinterwordspacing}{\spaceskip=\fontdimen2\font plus
\BIBentryALTinterwordstretchfactor\fontdimen3\font minus
  \fontdimen4\font\relax}
\providecommand{\BIBforeignlanguage}[2]{{%
\expandafter\ifx\csname l@#1\endcsname\relax
\typeout{** WARNING: IEEEtran.bst: No hyphenation pattern has been}%
\typeout{** loaded for the language `#1'. Using the pattern for}%
\typeout{** the default language instead.}%
\else
\language=\csname l@#1\endcsname
\fi
#2}}
\providecommand{\BIBdecl}{\relax}
\BIBdecl

\bibitem{8952223}
F.~{Recoules}, S.~{Bardin}, R.~{Bonichon}, L.~{Mounier}, and M.~{Potet}, ``Get
  rid of inline assembly through verification-oriented lifting,'' in \emph{34th
  IEEE/ACM International Conference on Automated Software Engineering
  (ASE'19)}.\hskip 1em plus 0.5em minus 0.4em\relax IEEE, 2019.

\bibitem{Rigger:2018:AXI:3186411.3186418}
M.~Rigger, S.~Marr, S.~Kell, D.~Leopoldseder, and H.~M\"{o}ssenb\"{o}ck, ``An
  analysis of x86-64 inline assembly in c programs,'' in \emph{Proceedings of
  the 14th ACM SIGPLAN/SIGOPS International Conference on Virtual Execution
  Environments (VEE'18)}.\hskip 1em plus 0.5em minus 0.4em\relax ACM, 2018.

\bibitem{1_STANNARD_2018}
\BIBentryALTinterwordspacing
O.~Stannard, ``[llvm-dev] [rfc] checking inline assembly for validity,''
  November 2018. [Online]. Available:
  \url{http://lists.llvm.org/pipermail/llvm-dev/2018-November/127968.html}
\BIBentrySTDinterwordspacing

\bibitem{gccdoc}
\BIBentryALTinterwordspacing
GCC, ``Extended asm - assembler instructions with c expression operands,''
  2020. [Online]. Available:
  \url{https://gcc.gnu.org/onlinedocs/gcc/Extended-Asm.html}
\BIBentrySTDinterwordspacing

\bibitem{Fehnker08}
A.~Fehnker, R.~Huuck, F.~Rauch, and S.~Seefried, ``Some assembly required -
  program analysis of embedded system code,'' in \emph{Eighth IEEE
  International Working Conference on Source Code Analysis and Manipulation
  (SCAM'08)}, 2008.

\bibitem{DBLP:conf/uss/CorteggianiCF18}
N.~Corteggiani, G.~Camurati, and A.~Francillon, ``Inception: System-wide
  security testing of real-world embedded systems software,'' in \emph{27th
  {USENIX} Security Symposium}.\hskip 1em plus 0.5em minus 0.4em\relax {USENIX}
  Association, 2018.

\bibitem{rustina_in_a_nutshell_2020}
\BIBentryALTinterwordspacing
F.~{Recoules}, S.~{Bardin}, R.~{Bonichon}, M.~{Lemerre}, L.~{Mounier}, and
  M.~{Potet}, ``\textsc{RUSTInA} in a nutshell,'' 2021. [Online]. Available:
  \url{https://binsec.github.io/new/publication/1970/01/01/nutshell-icse-21.html}
\BIBentrySTDinterwordspacing

\bibitem{10.1145/512927.512945}
G.~A. Kildall, ``A unified approach to global program optimization,'' in
  \emph{Proceedings of the 1st Annual ACM SIGACT-SIGPLAN Symposium on
  Principles of Programming Languages, {POPL’73}}.\hskip 1em plus 0.5em minus
  0.4em\relax ACM, 1973.

\bibitem{30527_gcc_bugzilla_2007}
\BIBentryALTinterwordspacing
D.~McCall, ``Use of input/output operands in \_\_asm\_\_ templates not fully
  documented,'' 2007. [Online]. Available:
  \url{https://gcc.gnu.org/bugzilla/show_bug.cgi?id=30527}
\BIBentrySTDinterwordspacing

\bibitem{Brumley2011}
D.~Brumley, I.~Jager, T.~Avgerinos, and E.~J. Schwartz, ``{BAP: A Binary
  Analysis Platform},'' in \emph{23rd International Conference on Computer
  Aided Verification (CAV 2011)}.\hskip 1em plus 0.5em minus 0.4em\relax
  Springer, 2011.

\bibitem{DBLP:conf/cav/BardinHLLTV11}
S.~Bardin, P.~Herrmann, J.~Leroux, O.~Ly, R.~Tabary, and A.~Vincent, ``The
  {BINCOA} framework for binary code analysis,'' in \emph{23rd International
  Conference on Computer Aided Verification ({CAV}'11)}.\hskip 1em plus 0.5em
  minus 0.4em\relax Springer, 2011.

\bibitem{Djoudi2015}
A.~Djoudi and S.~Bardin, ``Binsec: Binary code analysis with low-level
  regions,'' in \emph{Tools and Algorithms for the Construction and Analysis of
  Systems: 21st International Conference (TACAS'15)}.\hskip 1em plus 0.5em
  minus 0.4em\relax Springer, 2015.

\bibitem{DBLP:conf/wcre/DavidBTMFPM16}
R.~David, S.~Bardin, T.~D. Ta, L.~Mounier, J.~Feist, M.~Potet, and J.~Marion,
  ``{BINSEC/SE:} {A} dynamic symbolic execution toolkit for binary-level
  analysis,'' in \emph{{IEEE} 23rd International Conference on Software
  Analysis, Evolution, and Reengineering ({SANER}'16)}.\hskip 1em plus 0.5em
  minus 0.4em\relax {IEEE}, 2016.

\bibitem{DBLP:journals/fac/KirchnerKPSY15}
F.~Kirchner, N.~Kosmatov, V.~Prevosto, J.~Signoles, and B.~Yakobowski,
  ``Frama-c: {A} software analysis perspective,'' \emph{Formal Asp. Comput.},
  vol.~27, no.~3, 2015.

\bibitem{libatomic_ops6a7ee55}
\BIBentryALTinterwordspacing
I.~Maidanski, ``Fix compare\_double\_and\_swap\_double for clang/x86 in pic
  mode,'' September 2012. [Online]. Available:
  \url{https://github.com/ivmai/libatomic_ops/commit/64d81cd475b07c8a01b91a3be25e20eeca2d27ec}
\BIBentrySTDinterwordspacing

\bibitem{libatomic_ops04eeeb6}
\BIBentryALTinterwordspacing
------, ``Fix ao\_compare\_double\_and\_swap\_double\_full for gcc/x86 (pic
  mode),'' Mars 2012. [Online]. Available:
  \url{https://github.com/ivmai/libatomic_ops/commit/30cea1b9ea06c4c25cc219e1197dfac8dfa52083}
\BIBentrySTDinterwordspacing

\bibitem{libtomcrypt_cefff85}
\BIBentryALTinterwordspacing
P.~Pelletier, ``Add "memory" as a clobber for bswap inline assembly,''
  September 2011. [Online]. Available:
  \url{https://github.com/libtom/libtomcrypt/commit/cefff85550786ec869b39c0cb4a5904e88c84319}
\BIBentrySTDinterwordspacing

\bibitem{68095c4_gcc_bugzilla_2015}
\BIBentryALTinterwordspacing
S.~Boessenkool, ``Bug 68095 -- comment 4,'' 2015. [Online]. Available:
  \url{https://gcc.gnu.org/bugzilla/show_bug.cgi?id=68095#c4}
\BIBentrySTDinterwordspacing

\bibitem{10.1145/1984701.1984706}
C.~Parnin and C.~Treude, ``Measuring api documentation on the web,'' in
  \emph{Proceedings of the 2nd International Workshop on Web 2.0 for Software
  Engineering}.\hskip 1em plus 0.5em minus 0.4em\relax ACM, 2011.

\bibitem{10.1145/2901739.2901767}
D.~Yang, A.~Hussain, and C.~V. Lopes, ``From query to usable code: An analysis
  of stack overflow code snippets,'' in \emph{Proceedings of the 13th
  International Conference on Mining Software Repositories}.\hskip 1em plus
  0.5em minus 0.4em\relax ACM, 2016.

\bibitem{bardin_backward-bounded_2017}
S.~Bardin, R.~David, and J.~Marion, ``Backward-bounded {DSE:} targeting
  infeasibility questions on obfuscated codes,'' in \emph{{International
  Symposium on Security \& Privacy (S\&P'17)}}.\hskip 1em plus 0.5em minus
  0.4em\relax {IEEE}, 2017.

\bibitem{david_specification_2016}
R.~David, S.~Bardin, J.~Feist, L.~Mounier, M.~Potet, T.~D. Ta, and J.~Marion,
  ``Specification of concretization and symbolization policies in symbolic
  execution,'' in \emph{Proceedings of the 25th International Symposium on
  Software Testing and Analysis (ISSTA'16)}.\hskip 1em plus 0.5em minus
  0.4em\relax {ACM}, 2016.

\bibitem{DanielBR20}
L.~Daniel, S.~Bardin, and T.~Rezk, ``Binsec/rel: Efficient relational symbolic
  execution for constant-time at binary-level,'' in \emph{International
  Symposium on Security and Privacy (SP'20)}.\hskip 1em plus 0.5em minus
  0.4em\relax {IEEE}, 2020.

\bibitem{FeistMBDP16}
J.~Feist, L.~Mounier, S.~Bardin, R.~David, and M.~Potet, ``Finding the needle
  in the heap: combining static analysis and dynamic symbolic execution to
  trigger use-after-free,'' in \emph{Proceedings of the 6th Workshop on
  Software Security, Protection, and Reverse Engineering, SSPREW@ACSAC
  2016}.\hskip 1em plus 0.5em minus 0.4em\relax {ACM}, 2016.

\bibitem{DBLP-conf/kbse/KimFJJOLC17}
S.~Kim, M.~Faerevaag, M.~Jung, S.~Jung, D.~Oh, J.~Lee, and S.~K. Cha, ``Testing
  intermediate representations for binary analysis,'' in \emph{Proceedings of
  the 32nd {IEEE/ACM} International Conference on Automated Software
  Engineering ({ASE}'17)}.\hskip 1em plus 0.5em minus 0.4em\relax {IEEE}, 2017.

\bibitem{vsdoc}
\BIBentryALTinterwordspacing
Microsoft, ``Inline assembler,'' August 2018. [Online]. Available:
  \url{https://docs.microsoft.com/en-us/cpp/assembler/inline/inline-assembler?view=vs-2019}
\BIBentrySTDinterwordspacing

\bibitem{Maus2008}
S.~Maus, M.~Moskal, and W.~Schulte, ``Vx86: x86 assembler simulated in c
  powered by automated theorem proving,'' in \emph{12th International
  Conference on Algebraic Methodology and Software Technology
  (AMAST'08)}.\hskip 1em plus 0.5em minus 0.4em\relax Springer, 2008.

\bibitem{DBLP:phd/dnb/Maus11}
S.~Maus, ``Verification of hypervisor subroutines written in assembler,'' Ph.D.
  dissertation, University of Freiburg, Germany, 2011.

\bibitem{Schmaltz2012}
S.~Schmaltz and A.~Shadrin, ``Integrated semantics of intermediate-language c
  and macro-assembler for pervasive formal verification of operating systems
  and hypervisors from verisoftxt,'' in \emph{4th International Conference on
  Verified Software: Theories, Tools, Experiments (VSTTE'12)}.\hskip 1em plus
  0.5em minus 0.4em\relax Springer, 2012.

\bibitem{DBLP:journals/toplas/BalakrishnanR10}
G.~Balakrishnan and T.~W. Reps, ``{WYSINWYX:} what you see is not what you
  execute,'' \emph{{ACM} Trans. Program. Lang. Syst.}, vol.~32, no.~6, 2010.

\bibitem{DBLP:conf/fm/DjoudiBG16}
A.~Djoudi, S.~Bardin, and {\'{E}}.~Goubault, ``Recovering high-level conditions
  from binary programs,'' in \emph{{FM} 2016: Formal Methods - 21st
  International Symposium}.\hskip 1em plus 0.5em minus 0.4em\relax Springer,
  2016.

\bibitem{DBLP:conf/vmcai/BardinHV11}
S.~Bardin, P.~Herrmann, and F.~V{\'{e}}drine, ``Refinement-based {CFG}
  reconstruction from unstructured programs,'' in \emph{12th International
  Conference on Verification, Model Checking, and Abstract Interpretation
  ({VMCAI}'11)}.\hskip 1em plus 0.5em minus 0.4em\relax Springer, 2011.

\bibitem{DBLP-conf/vmcai/KinderK12}
J.~Kinder and D.~Kravchenko, ``{Alternating Control Flow Reconstruction},'' in
  \emph{13th International Conference on Verification, Model Checking, and
  Abstract Interpretation ({VMCAI}'12)}.\hskip 1em plus 0.5em minus 0.4em\relax
  Springer, 2012.

\end{thebibliography}

\end{document}